\documentclass[review]{elsarticle}

\usepackage{hyperref}

\journal{Journal of Systems and Software}
\usepackage{float} 
\usepackage{subcaption}
\usepackage{multicol}  
\usepackage{multirow}  
\usepackage{comment}
\usepackage{setspace}
\usepackage{verbatim}
\usepackage{xcolor}
\usepackage{soul}

\newcommand{\BGnote}[1]{{\color{blue}\textbf{[#1 (BG)]}}}

\newcommand{\DAnote}[1]{{\textbf{XXX #1 XXX}}}

\newcommand{\FFnote}[1]{{\color{orange}\textbf{[(FF) #1]}}}
\newcommand{\Hamta}[1]{{\color{red}\textbf{(Hamta) #1}}}
\newcommand{\RSnote}[1]{{\color{cyan}\textbf{(Roberto) #1}}}
\newcommand{\RStext}[1]{{\color{cyan}{#1}}}
\newcommand{\dropped}[1]{{\color{purple} \textbf{(dropped text)}}}
\newcommand{\cancel}[1]{{\setul{}{.2ex}\setstcolor{orange}\st{#1}\resetul}}

\newif\ifnocomments
\nocommentsfalse
\nocommentstrue  
\ifnocomments
    \renewcommand{\BGnote}[1]{}
    \renewcommand{\DAnote}[1]{}
    \renewcommand{\FFnote}[1]{}
    \renewcommand{\Hamta}[1]{}
    \renewcommand{\RSnote}[1]{}
    \renewcommand{\dropped}[1]{}
    \renewcommand{\cancel}[1]{}

    \renewcommand{\RStext}[1]{{#1}}
    \renewcommand{\Hamta}[1]{{#1}}
\fi

\begin{document}

\begin{frontmatter}

\title{OSCAR-P and aMLLibrary: Profiling and Predicting the Performance of FaaS-based Applications in Computing Continua}

\author{Roberto Sala, Bruno Guindani, Enrico Galimberti, Federica Filippini, Hamta Sedghani, Danilo Ardagna$^a$}
\address[mymainaddress]{Politecnico di Milano, Italy \\
Email: name.surname@polimi.it}
\author{Sebasti{\'a}n Risco, Germ{\'a}n Molt{\'o}, Miguel Caballer$^b$}
\address[mymainaddress]{Instituto de Instrumentaci\'on para Imagen Molecular (I3M). Centro mixto CSIC - Universitat Polit\`ecnica de Val\`encia, Camino de Vera s/n, 46022, Valencia, Espa\~na\\
Email: gmolto@dsic.upv.es,\{serisgal,micafer\}@i3m.upv.es}

\begin{abstract}
This paper proposes an automated framework for efficient application profiling and training of Machine Learning (ML) performance models, composed of two parts: OSCAR-P and aMLLibrary.
OSCAR-P is an auto-profiling tool designed to automatically test serverless application workflows running on multiple hardware and node combinations in cloud and edge environments.
OSCAR-P obtains relevant profiling information on the execution time of the individual application components.
These data are later used by aMLLibrary to train ML-based performance models.
This makes it possible to predict the performance of applications on unseen configurations.
We test our framework on clusters with different architectures (x86 and arm64) and workloads, considering multi-component use-case applications.
This extensive experimental campaign proves the efficiency of OSCAR-P and aMLLibrary, significantly reducing the time needed for the application profiling, data collection, and data processing.
The preliminary results obtained on the ML performance models accuracy show a Mean Absolute Percentage Error lower than 30\% in all the considered scenarios.
\end{abstract}


\end{frontmatter}


\section{Introduction}\label{sec:Introduction}
\vspace{-5pt}

Cloud computing is widely adopted today and has been used for years as the standard computing paradigm for enterprise-level distributed systems \cite{Vu2020}.
Despite its significant advantages in terms of costs and computational power accessibility, it is associated with significant challenges since data offloading leads to potential delays and increased expenses.
Therefore, it falls short in meeting the demands of contemporary applications, often Artificial Intelligence (AI)-based and associated with strict or almost real-time processing constraints.
Cloud users exhibit high sensitivity to delays and fluctuations, and they benefit from a newly-emerging paradigm called edge computing, striving to relocate applications closer to the point of data generation ~\cite{satyanarayanan2017emergence}.
This approach offers several advantages:
i) lower latency: by moving part of the computation where the data resides, we remove the round-trip-time delays needed to access the remote cloud data centers, resulting in faster response times and better performance; ii) reduced bandwidth usage: local processing at the edge removes the need to send vast amounts of data through the network, saving bandwidth and avoiding bottlenecks; iii) improved privacy: data processed on edge devices can be anonymized on the spot before communication, ensuring that potentially sensitive information is never shared with a central server; iv) better scalability: edge devices are usually cheap, and adding more to help with data processing is easy to implement and economical.
However, edge resources cannot be seen as a possible replacement for cloud computing since they are usually characterized by lower computational capacity and consequently become a bottleneck in the computation.
An integrated edge-cloud computing continuum enabling application components with different demands to be executed on the most appropriate resources is crucial to supporting complex application workflows effectively.

Together with the introduction of the computing continuum paradigm, another significant aspect characterizing the computational landscape in recent years is represented by the quick rise in popularity~\cite{das2018edgebench}\cite{baldini2017serverless} of the Function as a Service (FaaS) model.
It breaks down complex applications in workflows of small, usually short-lived components, which run on reusable services consisting of stateless containers activated by suitable events (e.g., a file upload).
Using containers instead of virtual machines (VMs) reduces both the development/deployment complexity and the resource usage.
Containers can be dynamically created or destroyed in response to workload variations, and their stateless nature allows them to be reused to serve another event as soon as the previous computation completes.
This increases the flexibility and makes the FaaS model suitable for scenarios characterized by light average workload interleaved with activity peaks.
Finally, the FaaS model in public clouds is characterized by per-millisecond costs bound to the actual resource usage~\cite{albuquerque2017function}.

Despite the benefits associated with distributed computing and FaaS models, evaluating the performance of a complex application whose components may be allocated at different levels of the computing continuum poses significant challenges.
Automated tools are needed to support the components profiling, measuring their execution times while exploiting variable resources and hardware configurations.
Furthermore, accurately predicting the performance of a given application at a target configuration is the key to proper planning and runtime management of the available resources.
As mentioned, edge devices with limited capacity often become the system bottleneck in case of workload spikes, and actions need to be taken to meet the performance objectives (e.g., limiting application execution time with a fixed threshold).

This paper proposes an integrated framework for the profiling and performance modeling of FaaS-based applications running in computing continua.
In our setting, the application execution is supported by OSCAR~\cite{Perez2019}, a state-of-the-art runtime environment that aims to create a highly parallel, event-driven, pipelined serverless environment to execute general-purpose data-processing computing applications.
This includes the usage of AWS Lambda to execute event-driven Docker-based applications.
On top of OSCAR, we developed \textit{OSCAR-P} (OSCAR-Profiler), a novel auto-profiling tool that can automatically test application workflows on different hardware and node combinations, gathering relevant information on the execution time of the individual components.
Moreover, OSCAR-P leverages \textit{aMLLibrary}~\cite{guindani2023amllibrary}, an open-source Machine Learning (ML) package we designed to automatically develop performance-predicting models for every service/resource pair, which can be combined to forecast the runtime of the entire workflow.
Compared to tools such as Kubeflow Katib~\cite{katib}, aMLLibrary does not require extensive configuration and deployment of computational resources, is more straightforward and portable, and supports utilities specific to performance modeling such as feature augmentation and selection.

The usage of ML for performance prediction is motivated by the ever-increasing complexity of modern software.
Often, the impact of input configurations and settings on software performance is not straightforward, preventing the use of analytical, white-box methods such as Petri nets \cite{barbierato2013modeling} and queuing networks \cite{gandini2014performance}.
Even in the simplest scenarios where accurate modeling is possible, the hypothesis and assumptions underlying analytical formulations prevent them from covering all use cases.
Therefore, an approach that does not require any knowledge of the internal details of the system, generally referred to as a black-box technique, is often preferred.
In particular, ML is the prominent category of black-box approaches for performance analysis \cite{didona2015using}.
The models built by aMLLibrary make it possible to predict the performance (e.g., the average execution time) of an application on unseen configurations with high accuracy.
This allows to limit the initial application profiling campaign since some configurations do not need to  be tested directly.
In general, this prediction capability enables efficient design-time decision making~\cite{SedghaniFA21} and runtime resource management \cite{filippini2023space4ai, li2019edge, kang2017neurosurgeon}, which are essential tasks for numerous cloud systems.

This paper extends our initial works in \cite{galimberti2023oscar} and \cite{guindani2023amllibrary}.
Compared to these papers, we i) discuss our contributions with a much larger level of detail, ii) add support for synchronous calls, partitioned applications, and AWS Lambda functions into the framework, and iii) extend our experimental campaign with four new applications representing different real-life use cases.

This work is organized as follows.
Section \ref{sec:RelatedWork} presents relevant literature works.
Section \ref{sec:oscar} provides a summary of the OSCAR framework and its architecture.
Section \ref{sec:oscar-p} thoroughly illustrates the OSCAR-P goals and its architecture, while Section \ref{sec:aMLLib} illustrates the capabilities of aMLLibrary and the performance analysis it supports.
Section \ref{sec:performance_models} describes the performance models used in our experiments.
Section \ref{sec:experiments} focuses on the experimental scenarios to validate our tools.
Conclusions and future works are discussed in Section \ref{sec:conclusions}.

\vspace{-8pt}
\section{Related work} \label{sec:RelatedWork}
\vspace{-5pt}

This section overviews recent works in technological fields relevant to this paper, namely benchmarking tools for FaaS and edge systems (Section \ref{subsec:rw-benchmarking}) and ML-based performance modeling (Section \ref{subsec:rw-perf-models}).


\vspace{-5pt}
\subsection{Benchmarking}  \label{subsec:rw-benchmarking}
As FaaS and edge computing rose to popularity, several cloud providers started offering their own FaaS platforms, each with different underlying technologies.
At the same time, many researchers tried to address the challenges of benchmarking applications deployed partially on the edge and partially on the cloud.  
SeBS (Serverless Benchmark Suite)~\cite{copik2021sebs} is a comprehensive benchmarking tool that systematically supports various applications, cloud resources, and commercial providers.  
EdgeBench~\cite{das2018edgebench} focuses instead on two providers, AWS IoT Greengrass and Microsoft Azure IoT Edge, using different performance metrics, and also compares the edge frameworks performance to the respective cloud-only implementations.
The framework in \cite{das2020performance} allows the user to specify the latency and cost requirements of application services, and it determines whether it is better to deploy them on the cloud or the edge.  
It is meant for applications on small smart edge devices, like smart cameras.  
DeFog~\cite{mcchesney2019defog} presents a benchmarking tool that tests an application across a cloud-only, edge-only, and cloud-edge environment by comparing performance across different deployments.  
The tool collects relevant metrics 
to understand how the application services can be better distributed across the computing continuum.
\cite{fuerst2023iluvatar} presents Ilúvatar, a low-latency FaaS control plane.  
Its design principles significantly reduce overhead compared to state-of-the-art control planes.
It introduces function-size-aware queueing policies to regulate worker load, improve latency, and balance utilization.
Finally, \cite{ustiugov2023enabling} proposes a methodology for generating synthetic traces from large-scale production workloads, focusing on Azure Functions across various scales and load factors.
The authors introduce In-Vitro, which uses an iterative approach to sample functions, minimizing the Wasserstein distance between the full trace and sampled trace distributions of essential workload characteristics.
Their experiments show that In-Vitro significantly enhances representativeness compared to random sampling methods.
Finally, \cite{FaaSRail} introduces FaaSRail, a tool that reconciles open-source workloads with real-world FaaS traces in production environments.
FaaSRail maps real traces to benchmarking suites, downscales invocations, emulates time variability, and replicates burstiness.

\vspace{-5pt}
\subsection{Performance models}  \label{subsec:rw-perf-models}
Many researchers focus on analyzing and predicting the performance of applications running on edge systems and, recently, the majority use ML models instead of analytical models.
For instance, \cite{disabato2021distributed} proposes some linear regression models to predict the execution time of Convolutional Neural Networks on edge devices, given constraints on memory and processing load.
\cite{kang2017neurosurgeon} focuses predominantly on mobile devices, specifically the cloud-only data-processing approach, and compares its efficiency to a partitioned approach where the computation is split between the cloud and mobile devices.  
In both~\cite{disabato2021distributed} and~\cite{kang2017neurosurgeon}, the ML performance models predict the application execution time by considering the number of floating point operations per second of the devices.
Authors of \cite{nawrocki2021cloud} employ several ML models alongside anomaly detection to properly configure a cloud-based Internet of Things (IoT) device manager while respecting Quality of Service (QoS) constraints.
Similarly, \cite{kirchoff2019preliminary} applies popular ML techniques to a workload prediction analysis on HTTP servers, showing that these techniques all achieve good predicting capabilities.

Performance modeling through ML is also applied to FaaS platforms.
SuanMing~\cite{grohmann2021suanming} is an integrated framework for learning performance degradation of microservice-based systems running in public and private clouds, using several  ML algorithms.  
Authors of \cite{mahmoudi2020temporal} propose the creation of a model to predict performance metrics by considering the application average response time for warm and cold requests, the requests arrival rate, and the system expiration threshold.
In \cite{mahmoudi2020performance}, the same authors propose a steady-state analytical performance model for improving the QoS of FaaS and reducing operating costs, trading off cost and performance by making the platform workload-aware.

Unlike previous works, OSCAR-P is focused on benchmarking the OSCAR framework, which can be deployed on top of any commercial and on-premises Cloud Management Platform (e.g., OpenStack) and hence has the critical benefit of being cloud-provider agnostic.
To the best of our knowledge, the literature proposes no other automated framework that integrates profiling with ML performance model building for the computing continuum.

\vspace{-8pt}
\section{The OSCAR Framework}\label{sec:oscar}
\vspace{-5pt}

\begin{figure}[ht]
    \centering
    \includegraphics[width=0.9\textwidth]{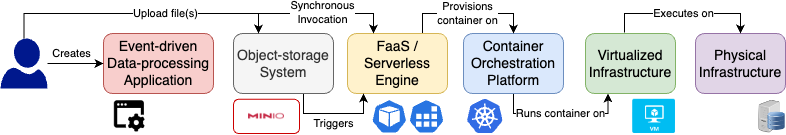}
    \caption{High-level architecture for event-driven container-based data-processing app.} \label{fig:oscar-HL-arch}
    \vspace{-5pt}
\end{figure}

OSCAR\footnote{\scriptsize{\url{https://oscar.grycap.net}}} is an open-source framework to quickly and efficiently support event-driven, data-processing, serverless applications packaged as Docker images along the computing continuum~\cite{risco2021serverless} (see Figure \ref{fig:oscar-HL-arch}).
They are executed in elastic Kubernetes (K8s) clusters that can be dynamically provisioned across multiple cloud back-ends thanks to the Infrastructure Manager\footnote{\scriptsize{\url{http://www.grycap.upv.es/im}}} (IM), an open-source tool that automatically creates and manages virtual infrastructures. 
Horizontal elasticity is provided by CLUES\footnote{\scriptsize{\url{http://github.com/grycap/clues}}},  
a management system that provisions and terminates the worker nodes to accommodate the cluster workload.
While one could choose a different container orchestration tool like Docker Swarm\footnote{\scriptsize\url{https://docs.docker.com/engine/swarm/}}, Kubernetes is preferred due to its greater scalability, richer ecosystem, and more robust features for complex deployments, making it a more versatile and powerful platform.
OSCAR can also perform cloud bursting into SCAR \cite{perez2018serverless}, facilitating the execution of Docker-based applications in AWS Lambda.

\begin{figure}[ht]
    \centering
    \includegraphics[width=0.9\textwidth]{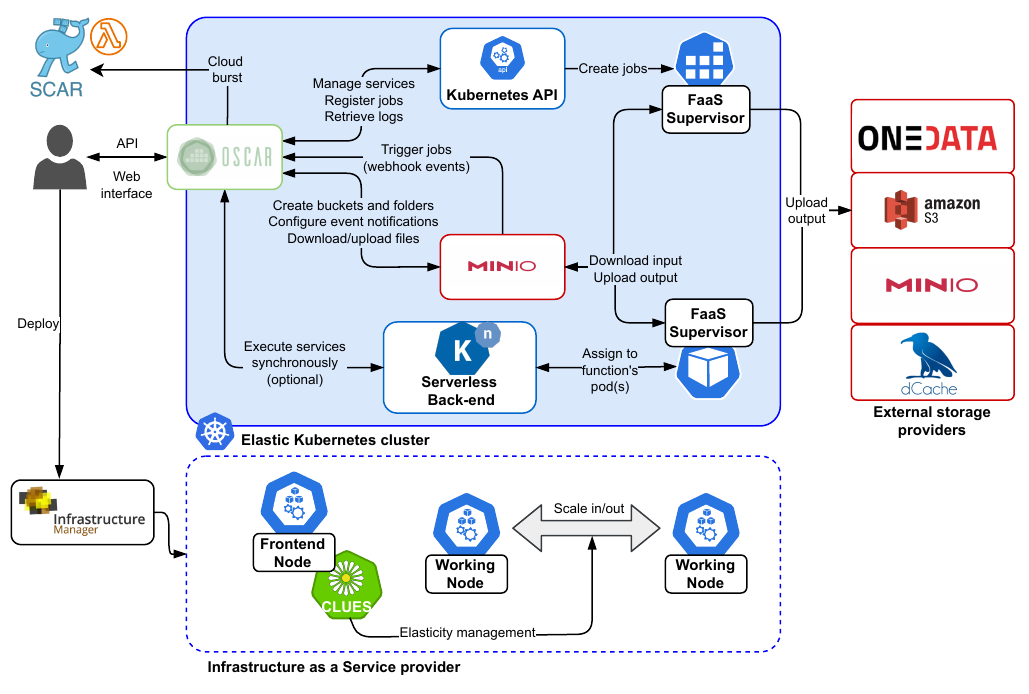}
    \caption{OSCAR architecture.} \label{fig:oscar-arch}
    \vspace{-5pt}
\end{figure}

The framework architecture is shown in Figure \ref{fig:oscar-arch}.
The following components are deployed inside the K8s cluster to support the OSCAR platform: i) MinIO\footnote{\scriptsize{\url{http://minio.io}}}, a high-performance multi-cloud object storage server 
; ii) Knative\footnote{\scriptsize{\url{https://knative.dev/docs/}}}, an open-source enterprise-level solution to build serverless and event-driven applications, used to support synchronous invocations, and iii) OSCAR Manager, the main service that manages the integration of the separate components.
The supported storage providers are: i) MinIO, deployed either internally or externally to the cluster; ii) Amazon S3, AWS object storage service that provides scalability, data availability, security, and performance in the public cloud; iii) Onedata, the global data access solution for science used by the EGI Federated Cloud; and iv) dCache, a system for storing and retrieving vast amounts of data, distributed among a large number of server nodes. 
The cluster can be managed through its REST API, either via web interface or command line.

Applications are created as services on the OSCAR cluster by providing: i) a Docker image to be used, ii) the input and output buckets of each service, and iii) a shell script to be executed inside the container.
OSCAR can create services by writing and uploading a Function Definition Language (FDL) configuration file on the cluster.
Once configured, the execution of a service is automatically triggered by uploading a file into its input bucket.
The result is delivered into the output bucket, which may be the input of another service.
If so, that service is automatically triggered and either executed immediately (if possible) or added to the job queue. Uploading multiple files triggers the highest possible number of parallel invocations supported by the specific cluster; all remaining function calls are scheduled for execution as soon as the running services are complete.

Serverless systems typically suffer from \textit{cold start} (see analyses in Sections \ref{sec:mask_detection_application_sync} and \ref{sec:fibonacci_application}), where the first invocations incur increased latency due to the underlying capacity allocation.
This is mitigated in OSCAR in two ways: pre-fetching the Docker images into the K8s cluster nodes to rapidly deploy the containers and maintaining a pool of user-defined active containers so that synchronous invocations via Knative can execute quickly.
Synchronous calls are made exclusively on the first component, while subsequent components are triggered by the arrival of files in the corresponding input buckets.

When performing inference via a Deep Neural Network (DNN), a single component can be replaced by two or more equivalent components running sequentially.
This is particularly relevant for AI applications since we can partition DNNs to run on different resources, exploiting more efficient alternative deployments according to the workload conditions \cite{SedghaniFA21, filippini2023space4ai}.

Despite our work being based on OSCAR, OSCAR-P and the prediction techniques we propose apply to other serverless frameworks too.
Indeed, the K8s and Knative platforms leveraged by OSCAR are widely used for FaaS both on-premises and in the cloud (see, e.g., OpenFaaS and Fission).
Thus, our approach can extend to any framework with compatible underlying technology.

\vspace{-8pt}
\section{OSCAR-P}\label{sec:oscar-p}
\vspace{-5pt}

This section describes OSCAR-P, the OSCAR Profiler that provides, alongside aMLLibrary, the novel contribution of this paper. 
OSCAR-P is built around OSCAR and its components (like K8s and MinIO) and acts as a director, configuring and coordinating the profiling activities and the data collection. 
The profiler aims to simplify and automate testing specific OSCAR application workflows on different hardware configurations, gathering data to train ML models through the aMLLibrary.
These models perform predictions on the response time of OSCAR/SCAR workflow components.

OSCAR-P requires the following input information:
i) the description of the physical or virtual resources to test, including access keys to AWS services and the private keys of physical clusters; 
ii) the description of the application components, 
e.g., their Docker images and hardware requirements; 
iii) the application parameters, including the input files, the distribution of data uploads (for asynchronous calls) and HTTP requests (for synchronous ones), and the parallelism levels; 
iv) the settings for ML models training in the aMLLibrary (see Section \ref{sec:aMLLib}).  
All these inputs are encapsulated in YAML files, eliminating the necessity for user interaction with the code.
A template of the individual configuration files is accessible on Zenodo \cite{zenodo_link}.


In the following sections, we will refer to a specific cluster setting to run the components of an application as a \textit{deployment}. 
OSCAR-P efficiently tests individual components after running the full workflow once, to reduce the time needed to test all possible deployments.
\begin{figure}[ht]
    \centering
    \includegraphics[width=0.9\linewidth]{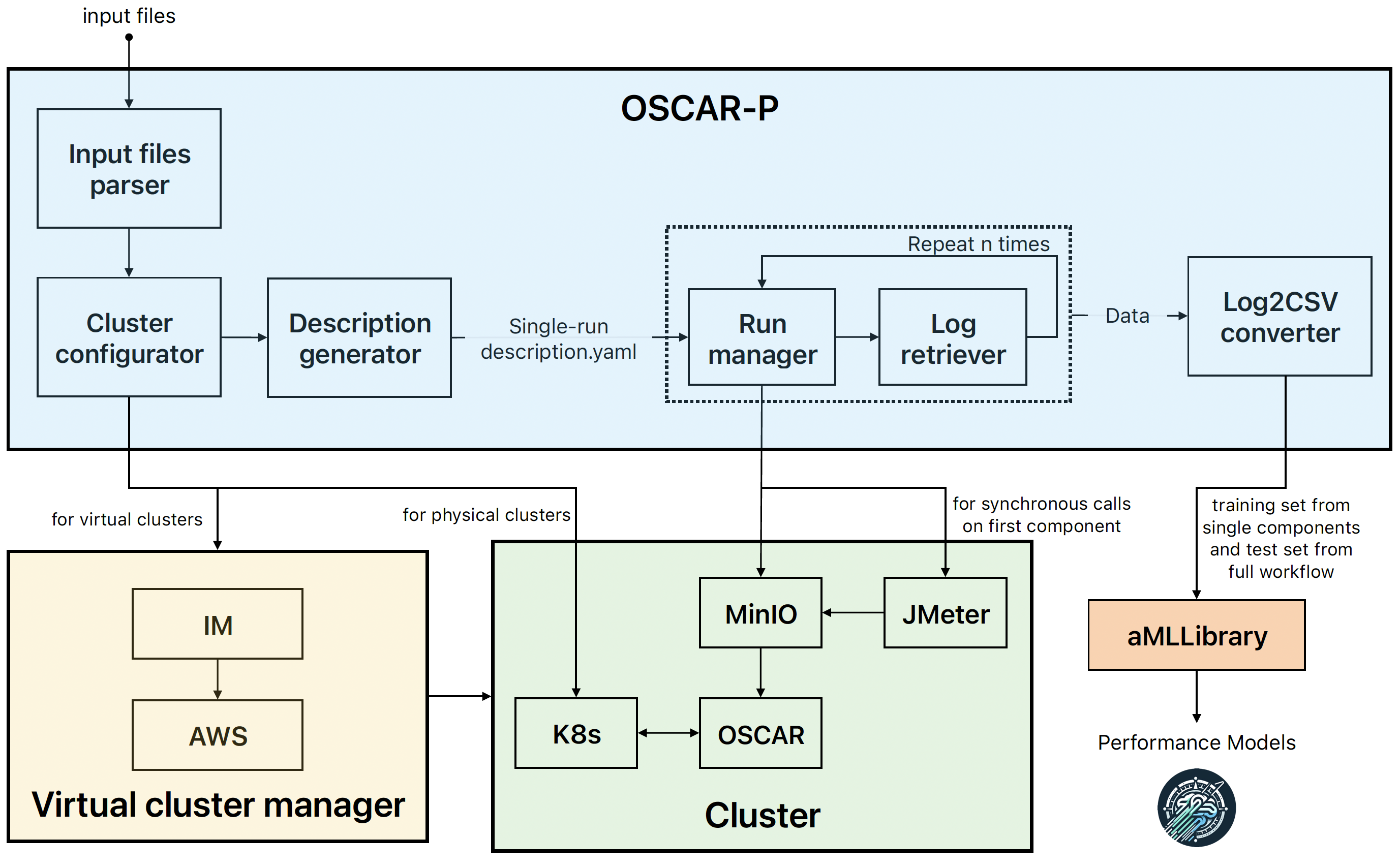}
    \caption{Profiling activity steps, OSCAR-P sub-components and interactions.}  \label{fig:oscarp-profiling-steps}
    \vspace{-5pt}
\end{figure}
Each OSCAR-P sub-component manages one step of the profiling activities, as illustrated in Figure \ref{fig:oscarp-profiling-steps}.

\vspace{-5pt}
\subsection{Input files parser}
Starting from the input files, OSCAR-P lists all the \textit{testing units}, i.e., the valid component-resource assignments within a deployment.
If a component is partitioned, all the partitions are considered part of the same testing unit.

\begin{figure}[ht]
    \centering
    \begin{minipage}[l]{.3\linewidth}
        \includegraphics[width=\linewidth]{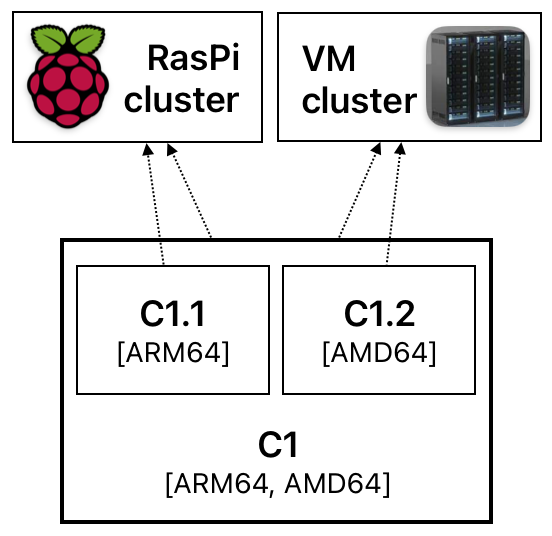}
        \caption{Example app.}\label{fig:simple-application-example}
    \end{minipage}
    \hfill
    \begin{minipage}[l]{.3\linewidth}
        \includegraphics[width=\linewidth]{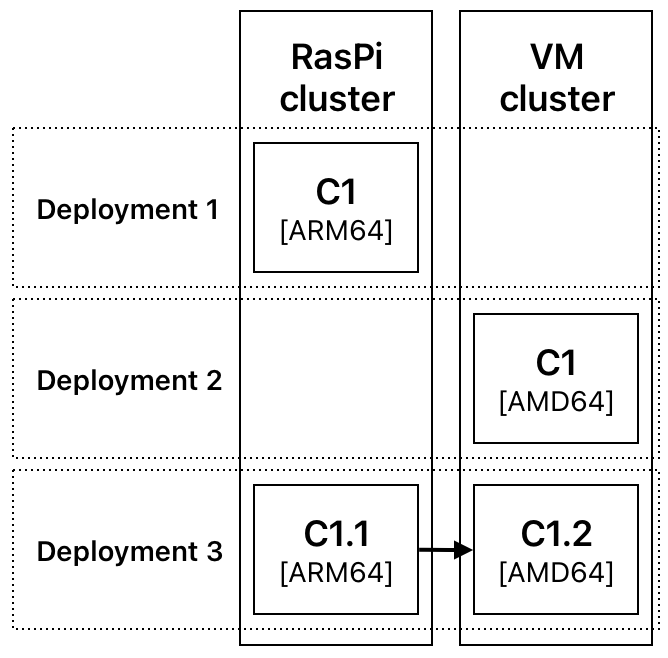}
        \caption{Testing units.}\label{fig:testing-units-example}
    \end{minipage}
    \hfill
    \begin{minipage}[l]{.3\linewidth}
        \includegraphics[width=\linewidth]{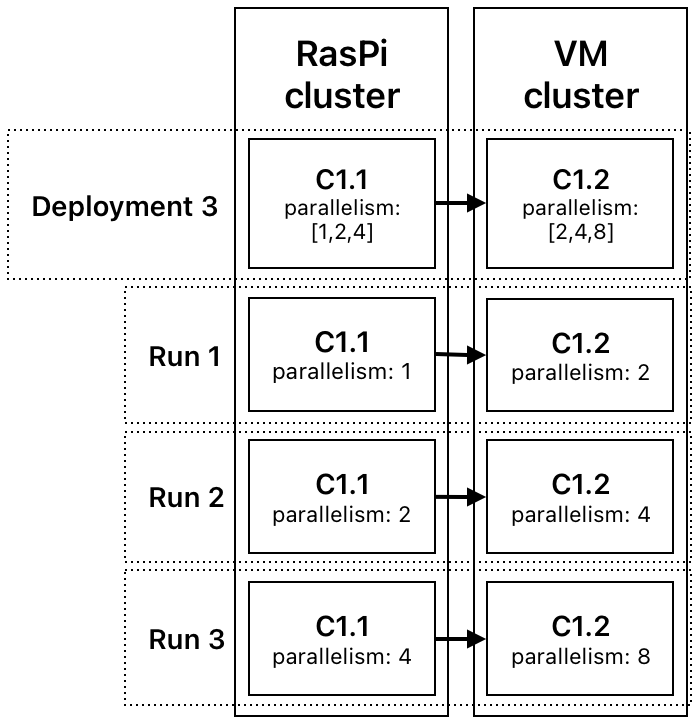}
        \caption{Deployment.}\label{fig:deployments-example}
    \end{minipage}
    \vspace{-5pt}
\end{figure}

As an example, illustrated in Figure \ref{fig:simple-application-example}, we consider an application including a single component (\textit{C1}) that can be deployed on a cluster of Raspberry Pi (arm64 architecture) or VMs (amd64 architecture).
The component can also be partitioned in \textit{C1.1} and \textit{C1.2}, each available for a single architecture (arm64 and amd64, respectively), generating the following testing units: i) C1 on the Raspberry Pi cluster; ii) C1 on the VM cluster; iii) C1.1 on the Raspberry Pi cluster and C1.2 on the VM cluster.
In this case, OSCAR-P creates a list of all the deployments to test, i.e., all the possible combinations of testing units of the different components (see Figure \ref{fig:testing-units-example}).
Each deployment contains a set of configurations, i.e., the cluster settings, but with different \textit{parallelism levels} for each component (i.e., the number of parallel instances executed).  
Figure \ref{fig:deployments-example} shows an example for a specific deployment.

\vspace{-5pt}
\subsection{Cluster configurator and description generation}
Before testing a deployment, the clusters need to be set up and correctly configured.
As shown in Figure \ref{fig:oscarp-profiling-steps}, OSCAR-P generates the necessary files for IM and K8s to instantiate both physical and virtual nodes.
During profiling, the cluster configuration undergoes periodic adjustments for subsequent runs, directly altering the number of worker nodes.
When OSCAR-P completes its runs, the cluster configurator mandates IM to destroy the virtual clusters.

Once the cluster configuration for a specific test is completed, OSCAR-P generates a descriptive YAML file for that deployment detailing all information needed to run the test.
This file contains the list of all the service requirements and a description of the clusters in use, 
and is updated with all the subsequent deployment runs, serving as a summary of the entire testing campaign.

\vspace{-5pt}
\subsection{Run manager}
At this stage, the description YAML file (mentioned in Figure \ref{fig:oscarp-profiling-steps}) is parsed to extract all the relevant information for the current run.
The OSCAR cluster is cleaned by removing all the remnants of past executions (if any) to ensure they do not interfere with the current execution.
Finally, OSCAR-P generates an FDL file with the information needed to build the new workflow (i.e., the required services and buckets) and applies it to the OSCAR cluster.
Since OSCAR (and FaaS in general) is event-driven, moving the required files in the input bucket of the first service triggers its execution and starts the run.
While for asynchronous calls the file upload is managed directly by the machine running OSCAR-P, for synchronous calls it is handled by JMeter\footnote{\scriptsize{\url{https://jmeter.apache.org}}}, an open-source software for load testing.
In particular, the JMeter client is automatically instantiated by IM before the run starts.
Given the input files and the desired load, JMeter can upload files either at a constant rate or by sampling arrival times from an exponential distribution.
Note that in the case of asynchronous calls, OSCAR-P input files are uploaded to a MinIO storage bucket and then moved to the input bucket.
This method allows to measure the processing time of an external request independently of the network connecting the end-user to the OSCAR cluster, something especially useful for simulating peak loads with multiple file uploads.
For the same reason, JMeter clusters are placed at the front-end component location for synchronous calls.

\begin{figure}[ht]
    \centering
    \includegraphics[width=.8\textwidth]{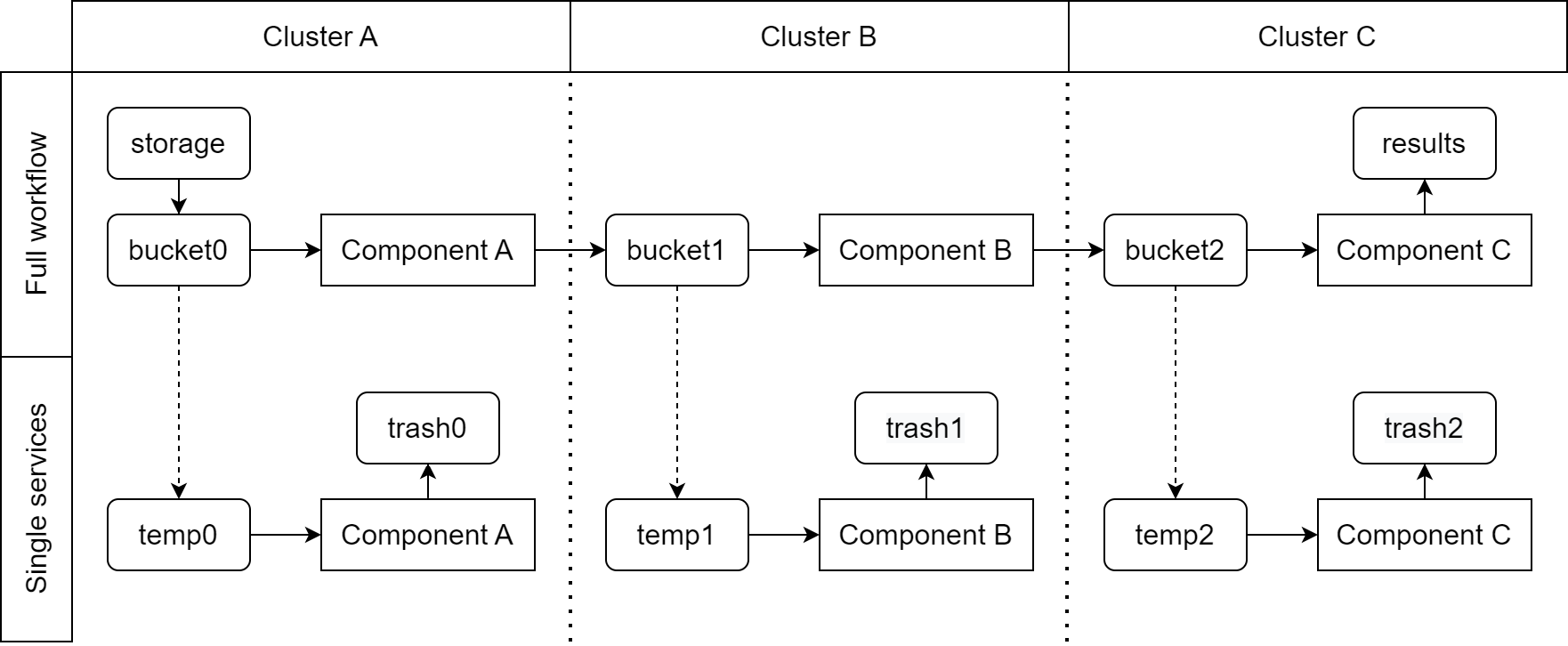}
    \caption{Full workflow and single component testing.}\label{fig:single-components-testing}
    \vspace{-5pt}
\end{figure}

When testing the full application workflow, once the execution of the first component is triggered, the process proceeds automatically, as the output of each component serves as the input for the following one.
On the other hand, single services are connected to temporary input/output buckets to control their execution, as depicted in Figure \ref{fig:single-components-testing}.

\vspace{-5pt}
\subsection{Log retriever}
Upon completion of a run, OSCAR-P collects and processes the relevant logs from OSCAR, Kubernetes (K8s), and potentially JMeter. These logs provide detailed information, including the scheduling time of a job  (i.e., a single component execution), the creation time of the corresponding \textit{pod} (the deployable computing unit in K8s), as well as the application's execution start and end times. This data is crucial for monitoring delays, wait times, and overheads.

\begin{figure}[ht]
    \centering
    \includegraphics[width=.9\textwidth]{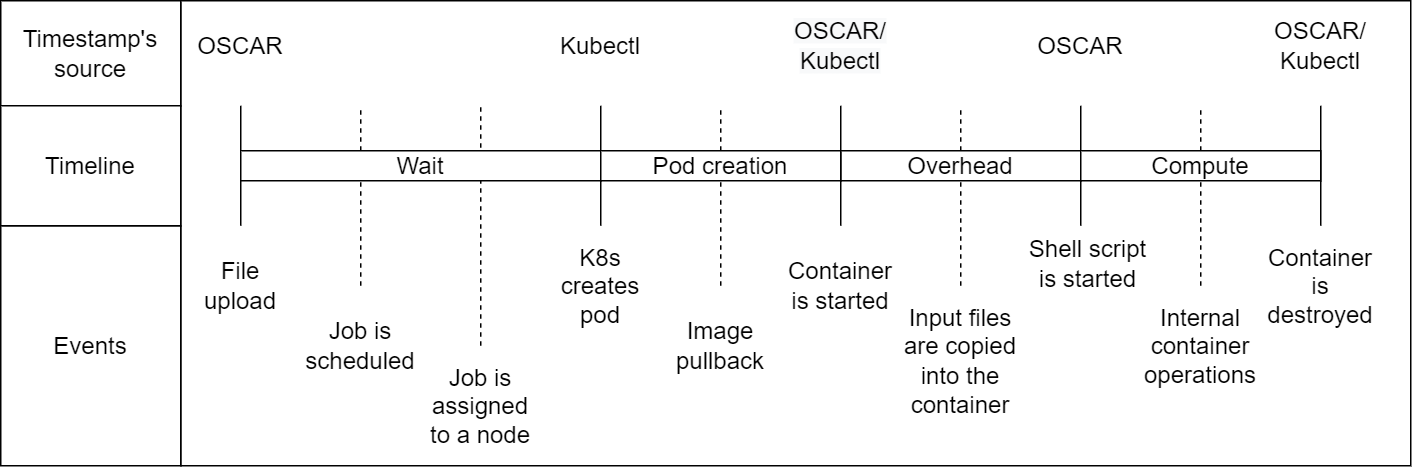}
    \caption{Timeline of a single job.}\label{fig:timeline-oscar}
    \vspace{-5pt}
\end{figure}

The collected data are organized in four time intervals, as reported in Figure \ref{fig:timeline-oscar}. The Network Time Protocol guarantees synchronization among the cluster nodes. 
The initial \textit{wait} interval starts when a file is uploaded in the input bucket.
The next interval corresponds to the \textit{pod creation} by K8s.
\textit{Overhead} starts when the pod is created and the underlying container is started, while \textit{compute} begins when the container log (obtained from OSCAR) reports that the application execution has already started.
This last interval ends when the job terminates, and the corresponding container is destroyed.

The \textit{Log2CSV-converter} OSCAR-P component merges the data collected for each tested configuration into a single CSV file, used by aMLLibrary to train a performance model for every service/resource pair.

\vspace{-8pt}
\section{aMLLibrary}\label{sec:aMLLib}
\vspace{-5pt}

aMLLibrary \cite{guindani2023amllibrary} is the last component in the OSCAR-P pipeline and builds the performance prediction models that are the final artifact produced by the framework.
aMLLibrary is an open-source, high-level Python package that is based on the scikit-learn toolkit\footnote{\scriptsize{\url{https://scikit-learn.org}}} and allows parallel training of multiple supervised ML models, supporting several pre-processing techniques, feature selection, and hyperparameter tuning.
It receives the profiling data collected by OSCAR-P and builds regression models to predict the performance of both the application individual components and the full workflow.

Overall, aMLLibrary implements an autoML solution, i.e., it trains multiple regression models and automatically selects the most accurate one.
By leveraging a highly accurate ML model, the performance of a specific application configuration can be estimated without direct observation, consequently reducing the size of the initial OSCAR-P profiling campaign.
Furthermore, the generalization capabilities of ML models allow to estimate the performance of a given application component for design-time decisions (e.g., how many nodes to allocate to each component of the OSCAR pipeline) \cite{SedghaniFA21} or runtime adaptive resource management (e.g., scaling nodes up and down according to runtime load variations) \cite{filippini2023space4ai}.
Also, when profiling some components of an application on a different hardware, we can can exploit the results from previous profiling campaigns for the unchanged components to reduce the overall profiling time.

As mentioned in Section~\ref{sec:Introduction}, general black-box approaches that do not require any knowledge of the internal details of the system, such as ML methods, are becoming popular in studying software performance because of the limitations of analytical, white-box models that often rely on unrealistic assumptions. 
Moreover, it would be impossible to formulate one single analytical model to cover all possible target applications.
Creating multiple individual, domain-specific models, each requiring thorough domain expertise, extensive effort, and significant profiling costs, is hardly affordable for large computing environments and applications. 
On the other hand, simulation-based approaches are characterized by excessive runtime evaluation costs, which makes it difficult to employ them for effective resource management decisions.

Within the OSCAR-P pipeline, the main strengths of the library are its ease of use, wide range of options, and robustness and efficiency of its training operations.
We discuss the first aspect in Section~\ref{subsec:aml-config-file} and the rest in Section~\ref{subsec:aml-features}.

\vspace{-5pt}
\subsection{Configuration files} \label{subsec:aml-config-file}
aMLLibrary is controlled by a simple text configuration file, given as input to the OSCAR-P framework (see \ref{app:aml-config} for a basic example of a configuration file).
\RStext{The file includes general settings for the campaign configuration, such as the ML models to build and the methods for hyperparameter selection and validation.}
It also lists the data pre-processing steps, if any.
The user can specify these settings at the beginning of the OSCAR-P profiling campaign in a concise and declarative way without the need to write any Python code.
An equivalent Python script would require hundreds of lines of code to achieve the same result (see the equivalent script to the configuration file in the Zenodo repository \cite{zenodo_link}).
The provided default settings for hyperparameter tuning are general enough to allow the tuning mechanism to find the appropriate parameter values, usually without requiring modifications by the user.

\vspace{-5pt}
\subsection{Features} \label{subsec:aml-features}
\begin{figure}[ht]
    \centering
    \includegraphics[width=\linewidth]{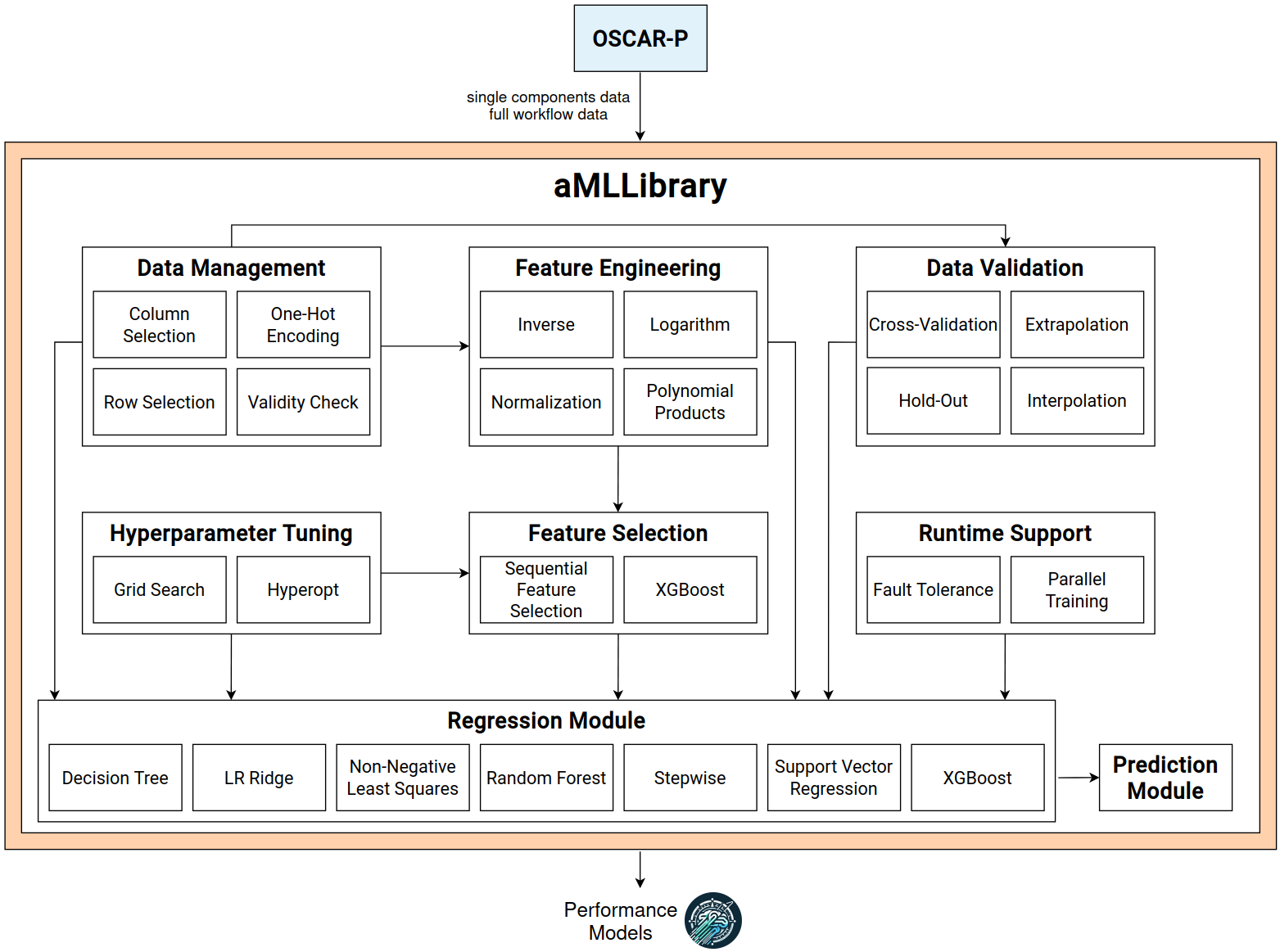}
    \caption{aMLLibrary components.}
    \label{fig:aml}
    \vspace{-5pt}
\end{figure}

We show the high-level architecture of aMLLibrary in Figure~\ref{fig:aml}.
The library has several useful perks for building performance models.
The parallel training of multiple models is supported: the user can specify the number of parallel cores to use, and the library automatically distributes the training experiments evenly among the parallel workers, even if the underlying scikit-learn models are limited to single-thread execution.
Furthermore, the library implements a fault tolerance mechanism by saving incremental progress checkpoints.

Users have complete control over the experimental campaign thanks to the many configuration options and flags available.
The library currently supports Decision Tree (DT), Non-Negative Least Squares (NNLS), Random Forest (RF), Ridge Linear Regression, Stepwise (a linear regression model which integrates the Draper-Smith feature selection technique \cite{draper1998applied}), Support Vector Regression (SVR), and XGBoost.
Hyperparameter tuning of these models can be performed either via grid search, by specifying the lists of values to be tested, or automatically via Bayesian Optimization (BO).
In the latter case, the integrated HyperOpt library\footnote{\scriptsize{\url{https://github.com/hyperopt/hyperopt}}} is used, and the user must provide prior probability distributions on the hyperparameters to optimize.

aMLLibrary includes plugins for several data pre-processing techniques, such as data normalization and one-hot encoding for discrete features, as well as other convenient tools, such as row selection and data validity checks.
It supports automatic feature engineering, e.g., computation of logarithms, inverse values, and feature products/polynomial expansion up to a given degree.
These tools can be helpful to unearth potentially relevant information hidden in the input features, such as quadratic dependencies and interaction terms.
Feature selection techniques are also supported, including forward Sequential Feature Selection (SFS) \cite{ferri1994comparative} and XGBoost importance weight selection.

\vspace{-5pt}
\subsection{Validation methods} \label{subsec:aml-validation-methods}
The user can choose among several validation methods to compute the Mean Absolute Percentage Error (MAPE) of a model, which is defined as $MAPE(y, \widehat{y}) = \frac{1}{N} \sum_{i=1}^N \left| \frac{y_i - \widehat y_i}{y_i} \right|$, where $y$ and $\widehat{y}$ are the vectors of true and predicted values, respectively.
In the performance evaluation literature~\cite{Lazowska1984}, MAPEs lower than 30\% for application execution times are usually considered enough to support runtime management decisions or capacity planning. 

Besides classical validation methods such as train-test splitting and Cross-Validation (CV), aMLLibrary offers methods such as interpolation and extrapolation, often used in ICT settings to build custom test sets \cite{maros2019machine}.
Here, interpolation means keeping some feature values within the feature space out of the training set and placing them in the test set to check the ability of the model to ``fill in the blanks'' of the feature space.
The goal of such validation is to verify whether it is possible to reduce the dimension of the training set, therefore conducting a profiling campaign with a coarser granularity (e.g., varying a cluster size by 8 cores instead of 2 or 4 cores).
Vice versa, extrapolation means keeping an entire area of the search space out of the training set to test the predicting capabilities of the model in an unexplored part of the feature space.
After the validation phase, aMLLibrary chooses the best model according to its validation MAPE and saves it to a binary file so that it can be used for further inference.
Finally, the library includes a prediction module that can be used to perform inference with an already-trained regression model.

\vspace{-8pt}
\section{Performance models}\label{sec:performance_models}
\vspace{-5pt}

This section presents the performance models used in our experimental scenarios, discussed in Section \ref{sec:experiments}.
Our general objective is to train ML performance models for individual OSCAR workflow services and to predict the performance of the full workflow.
Due to the substantial execution times and associated costs, we aim to avoid testing all possible combinations of a potentially complex application (see, e.g., ``recipe-transcriber'' in Section \ref{sec:recipe_transcriber_application}), where the number of experiments would scale exponentially with the number of components and resources to test. This prediction poses significant challenges, as the input rate of the subsequent component depends on the output rate of the previous component, further complicated by its configuration-specific nature. Additionally, we aim to generate ML models for individual components to facilitate reuse in case of setting changes, thereby avoiding the need to rerun the profiling campaign.

\begin{figure}[ht]
    \centering
    \includegraphics[width=.85\textwidth]{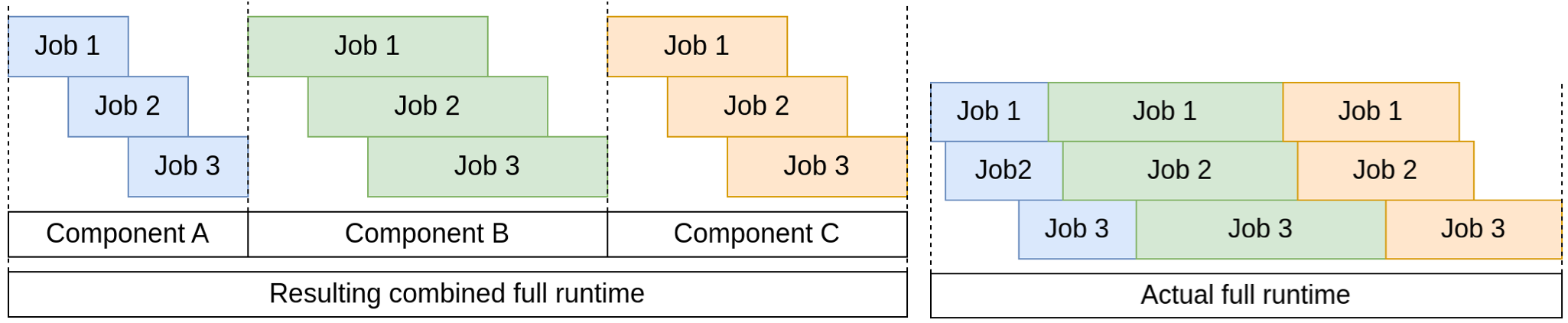}
    \caption{``Pipeline effect'' illustration.}\label{fig:pipeline-effect}
    \vspace{-5pt}
\end{figure}

In tests with asynchronous calls, we seek to estimate the total processing time for a collection of input files, considering that these files are processed all at once.
However, building a unified model encompassing all application components is not straightforward because of the \textit{pipeline effect}.
This effect arises during the execution of the full application, allowing subsequent components to start as soon as their predecessors have been completed. 
The performance models for single services do not capture this partial overlap (see Figure \ref{fig:pipeline-effect}), as they assume that a component only starts after all instances of the previous one have finished, \RStext{resulting in an overly conservative estimate of execution time.}
To mitigate this issue, we adjust the predictions for each component by subtracting the average execution time for a single job from the previous component.

\begin{figure}[ht]
    \centering
    \includegraphics[width=.6\textwidth]{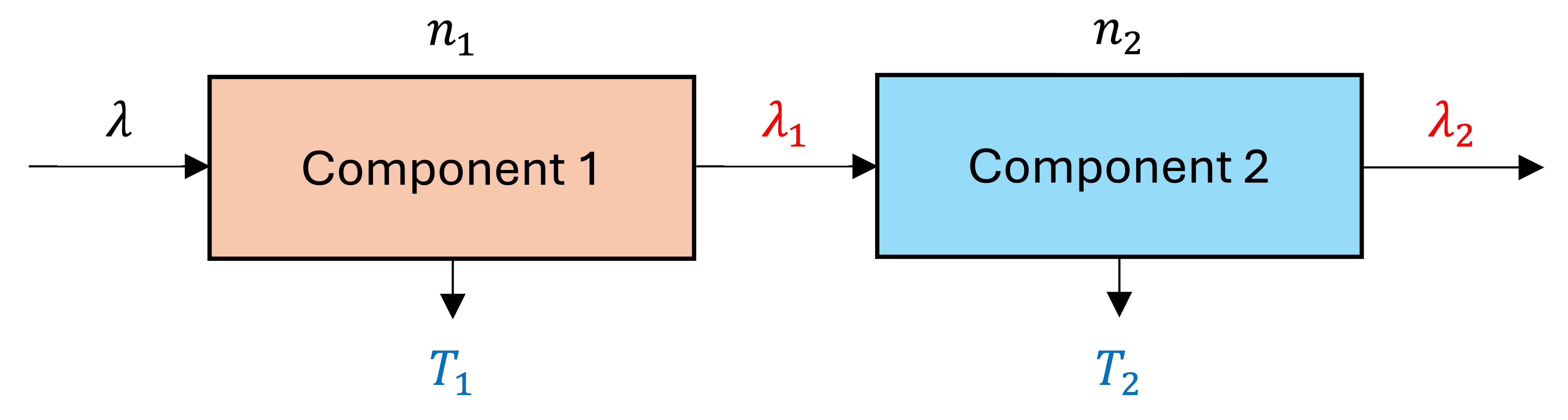}
    \caption{Overview of a test with synchronous calls on a two-component application, where the user-defined parameters for the specific run are $(\lambda, n_1, n_2)$.}\label{fig:synch tests pic}
\end{figure}

On the other hand, tests with synchronous calls aim to predict the execution time of individual inputs considering a continuous flow of requests with arrival rate $\lambda$.
The models for each component need to account for not only the number of parallel pods, but also the arrival rate of requests from the previous component.
In principle, the arrival rate of requests should match the value set by the user; however, this is not the case when one of the components saturates the available resources, which leads to a reduction in throughput compared to the initial rate.
Consider for example the scenario depicted in Figure \ref{fig:synch tests pic}, where the application includes two components.
If the first component saturates, the arrival rate of requests to the second component would be $\lambda_1<\lambda$.
Consequently, using the input arrival rate $\lambda$ to predict the performance of the second component would lead to overly conservative predictions for the full workflow.
To address this challenge, we generate models for all individual components to estimate their average execution times ($T_1$ and $T_2$ in the figure) and models to predict the throughput between one component and another (i.e.,  $\lambda_1$).

Note that this work focuses on the computational aspects of the cloud-edge continuum without addressing network communication delays.
This is equivalent to the assumption that the network is not a performance bottleneck.
Profiling and modeling its performance falls beyond the scope of our research.

\vspace{-10pt}
\section{Experimental analysis}\label{sec:experiments}
\vspace{-5pt}

This section overviews the experiments we perform with our framework, with the goal of demonstrating its usefulness in the collection of training data, post-hoc analysis of executions, and generation of highly accurate ML models.
We consider four example applications (Section \ref{sec:experimental-setup}), testing them across different clusters and configurations. 
Section \ref{sec:ML-models} overviews the ML techniques exploited for performance model generation and their hyperparameters.
Sections \ref{sec:mask_detection_application} through \ref{sec:fibonacci_application} illustrate the profiling setup and results for the four applications.
Finally, in Section \ref{subsec:exper-discussion}, we discuss the cost and time of profiling applications, showing the advantages of using OSCAR-P and aMLLibrary.
\RStext{The source code, input files, log data, and experimental results are available on Zenodo \cite{zenodo_link}.}

\vspace{-5pt}
\subsection{Target applications}\label{sec:experimental-setup}
The four applications presented in the following subsections represent three examples of AI workflows characterized by a variable number of heterogeneous components, and one simple Fibonacci application.
In particular, we consider object detection and video/audio transcription as representative AI tasks with different demands in terms of computational power.
We focus on AI applications because of the blooming interest in this field, and especially in the deployment of these applications at the edge.
Finally, the Fibonacci application differs from the previous cases in that the inputs are characterized by few bytes, allowing hundreds of requests per second.
Note that while we conduct numerous test campaigns to validate our framework for the first application, we conduct fewer experiments for the larger ones due to time and budget constraints.


\vspace{-5pt}
\subsubsection{Mask detection}\label{sec:mask-detector}
\begin{figure}[ht]
    \centering
    \includegraphics[width=0.95\textwidth]{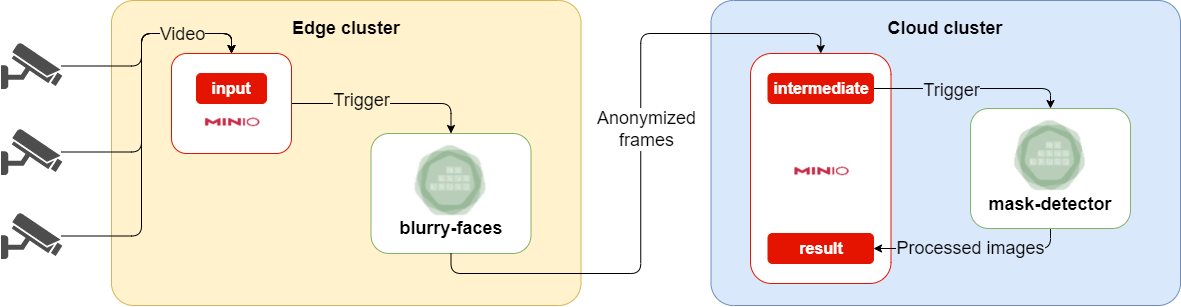}
    \caption{Mask detection workflow.}\label{fig-mask-detect}
    \vspace{-10pt}
\end{figure}

The first application we consider (see Figure \ref{fig-mask-detect}) 
was initially proposed in \cite{risco2021serverless} and consists of two components: \RStext{1) \textit{blurry-faces}, which receives a video as input, extracts a frame every 5 seconds via the \textit{FFmpeg} tool, and anonymizes it by blurring the faces of any person appearing in it by performing face recognition; 2) \textit{mask-detector}, which receives an image as input, detects the faces appearing in it, and decides whether or not they are wearing a mask.} 

The faces detection task is performed in both components via the YOLO (You Only Look Once) network, powered by the TensorFlow\footnote{\scriptsize{\url{https://www.tensorflow.org}}} object detection API and trained on the WIDER FACE dataset\footnote{\scriptsize{\url{http://shuoyang1213.me/WIDERFACE}}}.
This type of network is designed to detect multiple objects while examining each image only once, resulting in significantly faster performance while maintaining high accuracy.


By deploying the \textit{blurry-faces} component on the edge, we ensure the privacy of the people appearing in videos (because the server only sees anonymized frames) and reduce the latency associated with data transfer to the cloud.

\vspace{-5pt}
\subsubsection{Video searcher} \label{subsubsec:video-searcher-app}
\begin{figure}[ht]
    \centering
    \includegraphics[width=0.9\textwidth]{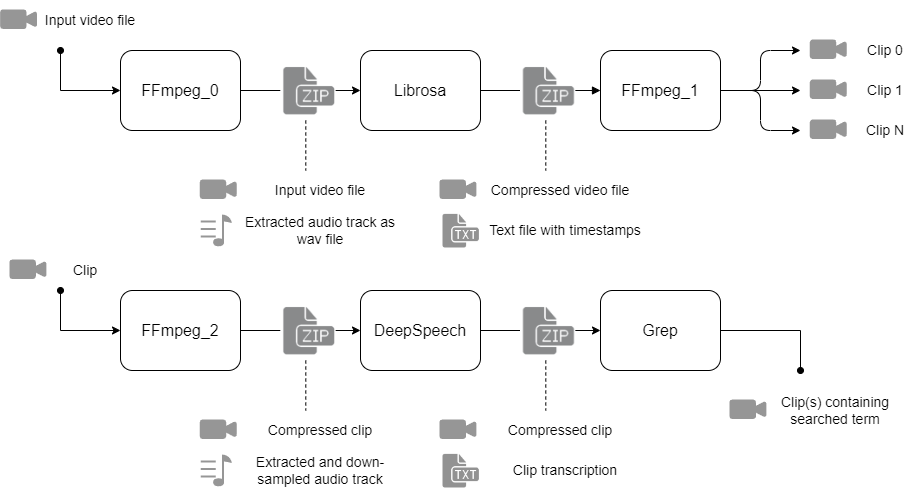}
    \caption{Video searcher workflow.}\label{fig:video-searcher-workflow}
    \vspace{-5pt}
\end{figure}

The second application (see Figure \ref{fig:video-searcher-workflow}) creates the transcription of a video, searches for one or more user-specified words in it, and returns a clip of the original video for every match. It includes six components:
\begin{enumerate}
    \item \textit{ffmpeg\_0} saves the audio track from the input video as a WAV file;
    
    \item \textit{Librosa} (based on the homonymous Python package\footnote{\scriptsize{\url{https://librosa.org}}}) analyses the audio track, looking for drops in the noise levels, which may indicate the end of a sentence.
    Each time the noise drops under a user-defined threshold, the component writes the corresponding timestamp in a text file;
    
    \item \textit{ffmpeg\_1} uses the timestamps produced by the previous component to cut the video into clips containing single sentences;
    
    \item \textit{ffmpeg\_2} extracts the audio track and lowers its sample rate to 16 kHz;
    
    \item \textit{DeepSpeech} transcribes the audio track of the clip using an open-source recurrent neural network implemented by Mozilla \cite{hannun2014deep};
    
    \item \textit{Grep} searches for the specified word(s) inside the transcription and copies the related video clip into the output bucket if it finds a match.
\end{enumerate}

\vspace{-5pt}
\subsubsection{Recipe transcriber} \label{subsubsec:recipe-transcriber-app}
The third application transcribes and recognizes the ingredients of a recipe, and it consists of seven components.
Components 1-5 are the same as in the video searcher application.
The last two are as follows: 

\begin{enumerate}
  \setcounter{enumi}{5}
  \item \textit{ffmpeg\_3} extracts a frame every 5 seconds from the original video;
  \item \textit{object-detector} identifies and localizes objects in a box.
\end{enumerate}

\vspace{-5pt}
\subsubsection{Fibonacci application} \label{subsubsec:fibonacci-app}
The last application (as in \cite{russo2024framework}) computes the $N$-th Fibonacci number where $N$ is specified in the input file.
In this case, the input file is only a few bytes, allowing exogenous loads up to hundreds of requests per second.
The application has a single component that performs an efficient, non-recursive computation of the requested Fibonacci number to avoid memory overheads.

\vspace{-5pt}
\subsection{ML models and hyperparameters settings}\label{sec:ML-models}

Once the profiling process is completed, aMLLibrary uses the collected data to build and compare the following ML models: XGBoost, Ridge Regression, Decision Tree, and Random Forest.
The hyperparameters are obtained via BO tuning (see Section \ref{sec:aMLLib}), fixing the maximum number of evaluated hyperparameter sets to 10 and considering the values or prior distributions listed in Table \ref{tab:oscarp-hyperparameters} reported in \ref{app:aml-hypers}.

For each experiment, we train two instances of each regression model, one involving feature selection (see Section \ref{sec:aMLLib}) and the other without it.
Each row of the training dataset provided by OSCAR-P represents an individual application component execution, while columns (also referred to as ``features'') identify the corresponding configuration characteristics in terms of, e.g., parallelism level. 
To cope with the execution time variability in practical settings, OSCAR-P profiles each configuration multiple times (three in our case), resulting in as many dataset rows with potentially different performance measurements.

As mentioned, the primary factor influencing performance prediction is the \textit{parallelism level}, which usually corresponds to the maximum number of available cores for all services (a relevant exception to this assumption will be discussed in Section \ref{sec:recipe_transcriber_application}).  
Since for all the applications and components we consider each pod requires exactly one core, the parallelism level directly translates to the maximum number of concurrently running jobs.
For simplicity, we refer to this feature as \textit{cores}.
Additionally, in tests with synchronous calls, the \textit{input workload}/\textit{throughput} is also a pivotal feature for accurate performance prediction.

Aside from \textit{cores}, we also use \textit{1/cores} and \textit{log(cores)} as features, as they are relevant quantities to predict the performance of parallel systems \cite{maros2019machine}. 
Moreover, we perform polynomial expansion of features up to the second degree.
We also test the interpolation and extrapolation capabilities of the models (see Section \ref{sec:aMLLib}).
As mentioned, the goal of interpolation tests is to understand how dense our profiling campaign must be to achieve good results and if we can still obtain accurate models with smaller datasets.
At the same time, extrapolation tests aim to understand the behavior of the models on unseen, expensive-to-evaluate configurations.
This means, e.g., analyzing whether it is possible to start from experiments run on a limited number of files and predict the performance for larger inputs.
We perform interpolation tests in all cases and extrapolation tests when they are significant (as discussed in the following sections).

The aMLLibrary models are generated on a server with two Xeon E5-2620 v2 processors, with 6 cores each, and 32 GB RAM.
More details on the composition of training sets in different tests are shown in the respective sections, along with the results tables.
Note that, in the worst-case scenario, the library trains all ML learning models within approximately ten minutes.


\vspace{-5pt}
\subsection{Results with the Mask detection application}\label{sec:mask_detection_application}
This section reports the profiling setup and the results obtained for \textit{Mask detection}.
We test the application both on physical nodes and cloud VMs, considering different workload injection models and evaluating the impact of neural network partitioning on the first component. 
In particular, Section \ref{sec:mask_detection_application_async} focuses on testing the performance of asynchronous calls, varying the components and resources configurations, the number of input videos, and the impact of different edge resources.
Section \ref{sec:mask_detection_application_sync} considers instead synchronous calls performed via JMeter, which allows to evaluate the prediction accuracy of our models when combining the impact of different resource configurations and throughputs.

\vspace{-5pt}
\subsubsection{Tests with asynchronous calls}  \label{sec:mask_detection_application_async}
In this first scenario, we consider asynchronous calls to the \textit{Mask detection} application, where the videos to process are uploaded to the input bucket of the first component in batches of variable size.



\begin{figure}[ht]
    \centering
    \subfloat[Full workflow]{
        \includegraphics[width=0.32\textwidth]{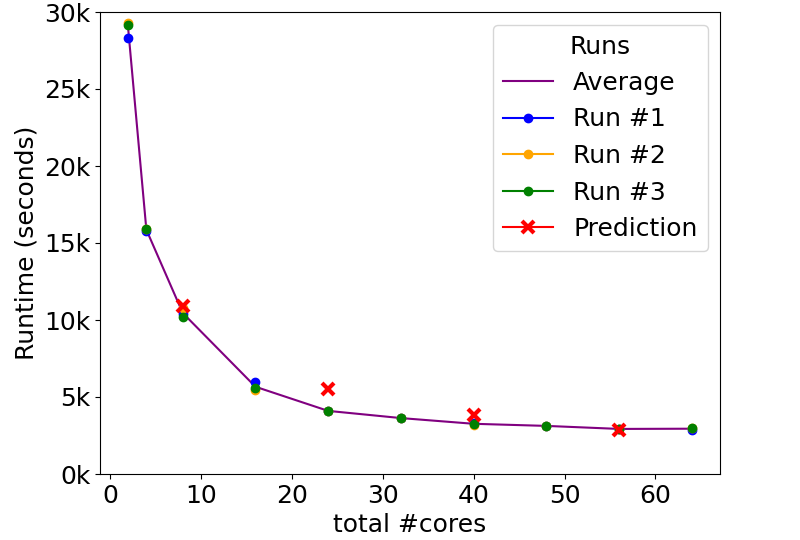}\label{fig:full_combined_15_mins}
    }
    \subfloat[\textit{blurry-faces} component]{
        \includegraphics[width=0.32\textwidth]{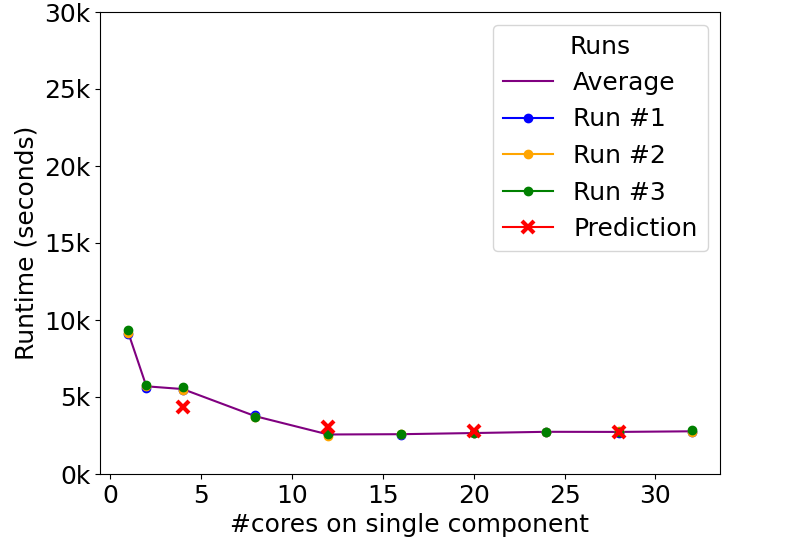}
    }
    \subfloat[\textit{mask-detector} component]{
        \includegraphics[width=0.32\textwidth]{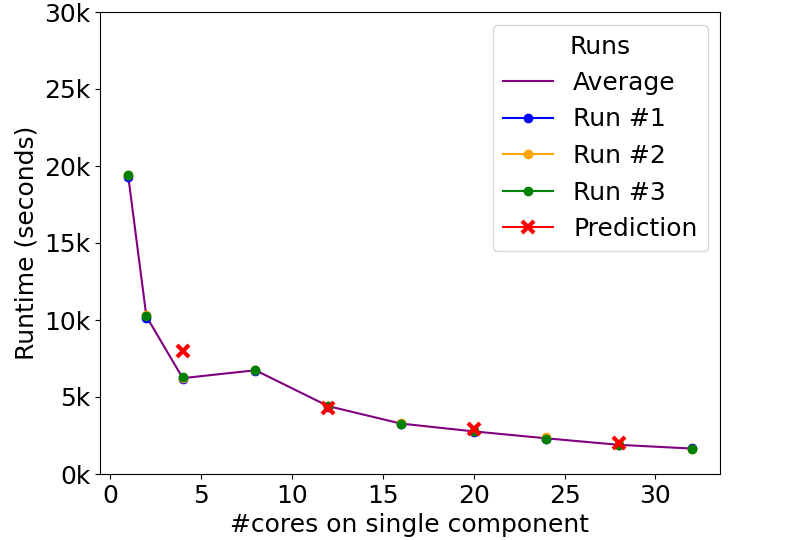}
    }
    \caption[]{Scenario 1: interpolation results.}\label{fig:interpolation_15_mins}
    \vspace{-5pt}
\end{figure}

\vspace{-5pt}
\paragraph{Scenario 1: impact of different parallelism levels}
In the first scenario, we consider a single batch of ten videos of 15 minutes each.
We analyze the impact of different configurations, running each component on up to 8 VMs with 4 cores and 8 GB of memory each, with a total of 32 cores, i.e., 32 parallel pods.
A runtime/core plot showcasing the impact of the number of cores on performance is reported in Figure \ref{fig:interpolation_15_mins}.
The runtime is the interval between the start of the first job and the completion of the last one, including wait times and overhead.

We use the collected data to generate prediction models for the runtime of the single components and of the entire application (see  Figure \ref{fig-mask-detect}).
Moreover, we investigate whether combining the predictions obtained for \textit{blurry-faces} and \textit{mask-detector} (what we call the \textit{combined model}) allows to predict the full workflow runtime with a reasonable level of accuracy.
We then analyze the interpolation capabilities of our models, generating a test set with execution times collected on 4, 12, 20, and 28 cores while using the remaining data for training.
Table \ref{table:MAPE-interpolation-extrapolation} collects the MAPE on the test set for all trained regression models; the best models (corresponding to values in bold) are used to generate the predictions reported in Figure \ref{fig:interpolation_15_mins}.
The results show that our best model, namely Ridge Regression without SFS, is able to predict the total runtime of the full workflow with an error of 14.73\%.


\begin{table}[h!t]
\centering
\resizebox{\linewidth}{!}{
\begin{tabular}{ |p{2.5cm}|p{1cm}|p{2.5cm}|p{2.5cm}|p{2.5cm}|p{2.5cm}|p{2.5cm}|p{2.5cm}| }
\cline{3-8}
\multicolumn{2}{c}{} & \multicolumn{3}{|c|}{MAPE interpolation, \textit{Scenario 1}} & \multicolumn{3}{|c|}{MAPE extrapolation, \textit{Scenario 2}} \\
\hline
Algorithm & SFS & Full workflow & \textit{blurry-faces} & \textit{mask-detector} & Full workflow & \textit{blurry-faces} & \textit{mask-detector} \\ \hline
\multirow{2}{*}{Ridge Regression} 
& Yes & 23.05 & 14.80 & 19.32 & \textbf{9.82} & \textbf{9.60} & 17.19  \\
\cline{2-8}
& No & \textbf{14.73} & 11.48 & \textbf{10.14} & 70.60 & 43.01 & 73.63 \\\hline
\multirow{2}{*}{Decision Tree} 
& Yes & 46.78 & 13.03 & 35.1 & 17.94 & 22.01 & 24.73 \\
\cline{2-8}
& No & 30.28 & 9.29 & 38.12 & 23.40 & 27.8 & 34.04 \\\hline
\multirow{2}{*}{XGBoost} 
& Yes & 46.99 & 13.12 & 39.48 & 17.23 & 17.6 & \textbf{10.09} \\
\cline{2-8}
& No & 46.81 & 12.94 & 39.48 & 17.39 & 21.3 & 24.37 \\\hline
\multirow{2}{*}{Random Forest} 
& Yes & 19.78 & \textbf{3.34} & 15.75 & 20.36 & 19.12 & 13.24 \\
\cline{2-8}
& No & 28.41 & 16.15 & 27.45 & 89.19 & 60.54 & 32.53 \\\hline
\end{tabular}
}
\caption{MAPE [\%] on interpolation over \textit{parallelism level} (\textit{Scenario 1}) and on extrapolation over \textit{batch size} (\textit{Scenario 2}).}\label{table:MAPE-interpolation-extrapolation}
\vspace{-5pt}
\end{table}

\vspace{-5pt}
\paragraph{Scenario 2: impact of different batch sizes}

\begin{figure}[ht]
    \centering
    \subfloat[Runtimes with 5, 10 and 15 files]{
        \includegraphics[width=0.4\textwidth]{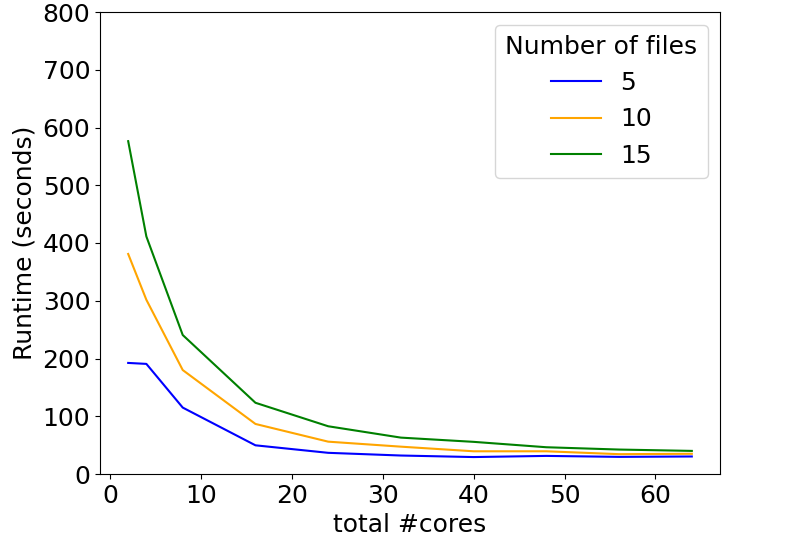}  \label{subfig:runtime_10_a}
    }
    \subfloat[Extrapolation results]{
        \includegraphics[width=0.4\textwidth]{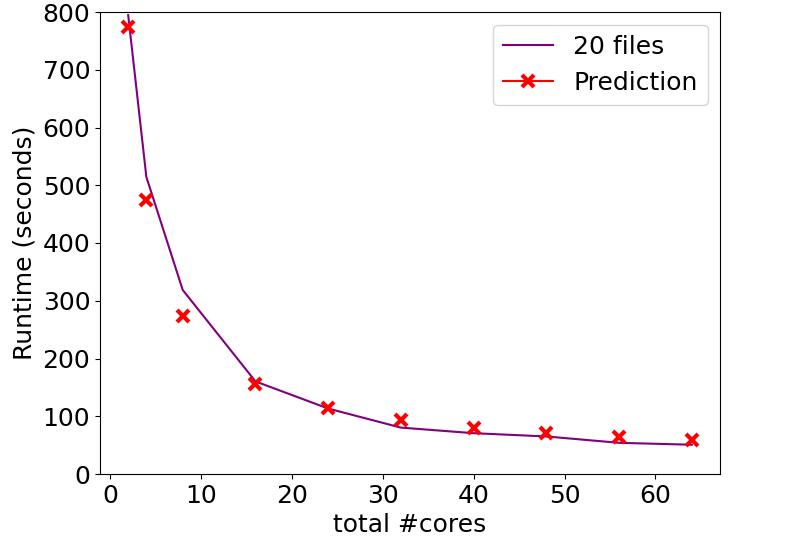}  \label{subfig:runtime_10_b}
    }
    \caption[]{Scenario 2: extrapolation results for the full workflow runtime.}\label{fig:runtime_core_full_10_secs}
    \vspace{-5pt}
\end{figure}

We test the application on the same configurations considered in Scenario 1, evaluating the combined impact of various parallelism levels and input batch sizes on the components runtime.
In particular, we consider 10-second-long videos uploaded in batches of 5, 10, 15, and 20.
We aim to generate and test performance prediction models with good extrapolation capabilities, i.e., to check whether a model trained on data collected with batches of size 5, 10, and 15 can accurately predict the components runtime when 20 files are submitted. 
%
We report MAPEs for the trained models in Table \ref{table:MAPE-interpolation-extrapolation}.
As in the previous scenario, we combine the two best models for the individual services to predict the whole application runtime, achieving a best MAPE of 9.82\%.
Figure \ref{fig:runtime_core_full_10_secs} reports the execution times for the full application with 5, 10, and 15 files as input (Figure \ref{subfig:runtime_10_a}), and the times for 20 files along with the corresponding extrapolation predictions (Figure \ref{subfig:runtime_10_b}).
%

\vspace{-5pt}
\paragraph{Scenario 3: evaluation on different edge resources}

We conduct the third campaign on a cluster of 3 Raspberry Pi devices, processing batches of ten videos lasting 10 seconds each and exclusively testing the \textit{blurry-faces} component.

Here, we aim to demonstrate the logging capability of OSCAR-P, as depicted in Figure \ref{fig:timeline-oscar}.
We assess the performance of the \textit{blurry-faces} component in two different configurations: on the Raspberry Pi cluster alone, and with the addition of an Intel Movidius Neural Compute Stick 2 (NCS2)\footnote{\scriptsize{\url{https://intel.com/content/www/us/en/developer/tools/neural-compute-stick/overview.html}}} USB accelerator.
We conduct all tests with an 8-core configuration.

To precisely analyze how this device affects the component performance, we measure the duration of different computation phases:
i) splitting the video into frames;
ii) overhead for loading Python modules;
iii) image reading from disk;
iv) model loading, either in the device RAM or on the Intel NCS2;
v) inference;
vi) image blurring;
vii) copying the results onto disk.
It is worth noting that since the first task is executed on CPUs, we should not expect any impact on its performance.
On the other hand, Figure \ref{fig:ncs-pi-bars} in \ref{app:exper-mask} highlights that the model computation on the stick is over 7 times faster compared to the CPU, while the model loading is only one second slower.

Figure \ref{fig:ncs-pi-lines} reports the \textit{blurry-faces} component runtime with and without the NCS2 while varying the number of input files.
The benefit of acceleration becomes more apparent as the number of files increases since they are processed sequentially.
However, despite the significantly faster computation with the NCS2, the overall performance gains are mitigated by the model load phase which is slightly slower on the stick compared to the device RAM.

\vspace{-5pt}
\subsubsection{Tests with synchronous calls}  \label{sec:mask_detection_application_sync}
In the fourth and fifth testing scenarios, we focus on synchronous calls to OSCAR services.
We inject an input workload $\lambda$ via JMeter, imposing a ramp-up of 10 seconds to avoid measuring runtimes during the warm-up period and testing different ranges depending on the number of cores.
For example, for component configurations including 2 cores, we consider $\lambda \in [0.05, 0.3]$ requests per second (req/s).
At 4 cores, since the system can sustain higher throughput values before reaching the saturation level, we test $\lambda \in [0.1, 0.6]$ req/s.
Each workload value is kept constant for 10 minutes.
Note that in each setup, half of the cores of the first component are warm to limit the impact of the cold start.
The synchronous scenarios are more critical because, as the load increases, the individual configurations saturate, resulting in larger variability in execution time.
This outcome is expected as it is consistent with results using other modeling techniques such as M/G/k queues \cite{kendall1953stochastic}.
Additionally, ML yields more effective results than the M/G/k approach as the latter requires expensive simulation time to estimate the components demand, whereas ML model predictions can be obtained in milliseconds.
%

\vspace{-5pt}
\paragraph{Scenario 4: impact of the number of cores and throughput values}

\begin{figure}[ht]
    \centering
    \subfloat[\textit{Full workflow}]{
        \includegraphics[width=0.32\textwidth]{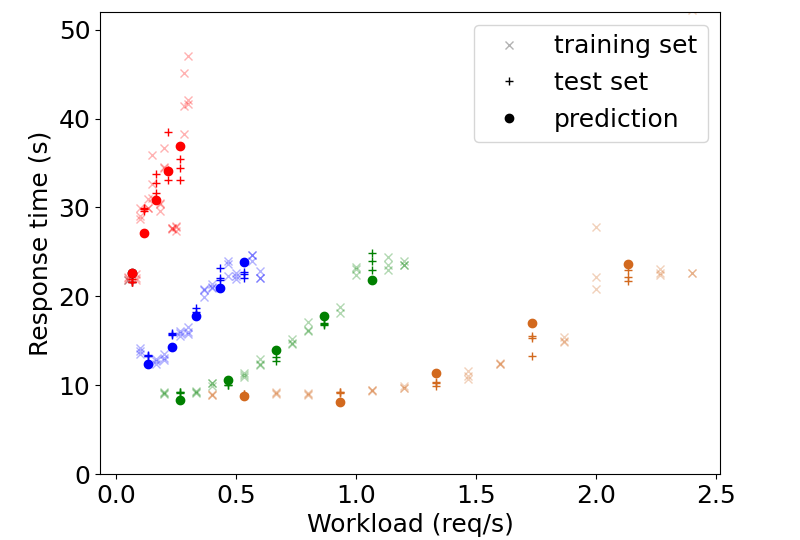}\label{pic_full_workflow_synchro}
    }
    \subfloat[\textit{blurry-faces} component]{
        \includegraphics[width=0.32\textwidth]{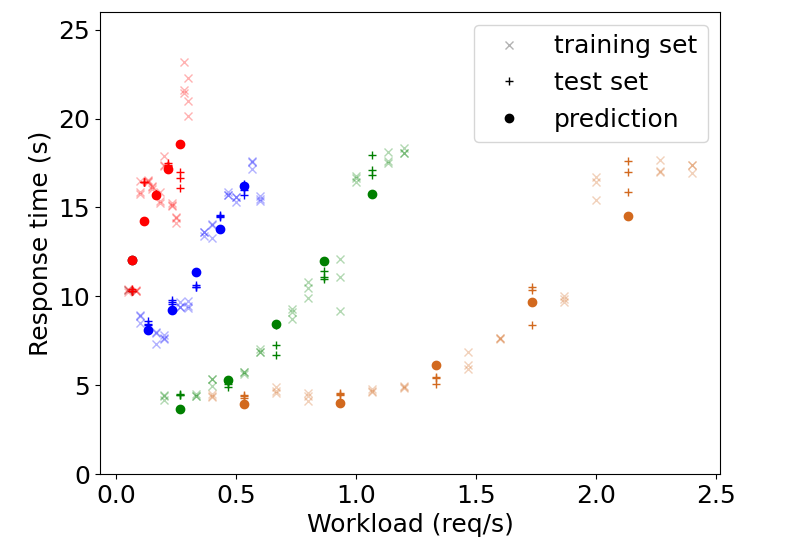}\label{pic_blur-faces_synchro}
    }
    \subfloat[\textit{mask-detector} component]{
        \includegraphics[width=0.32\textwidth]{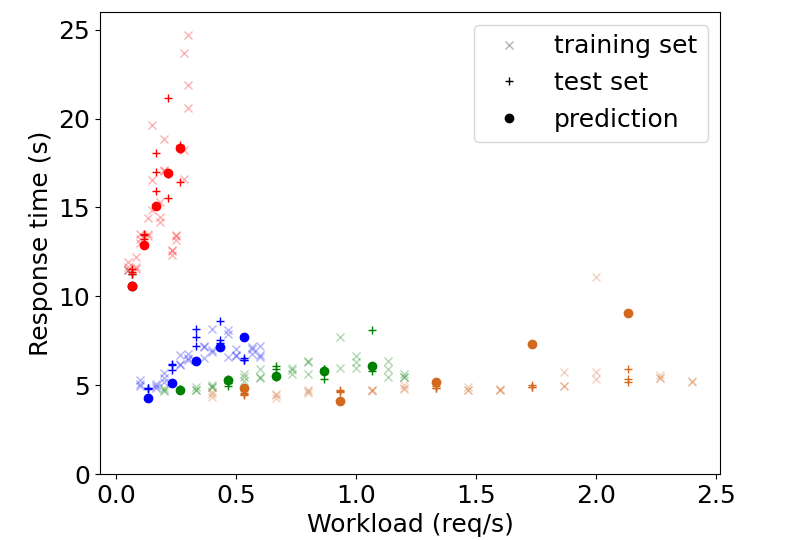}\label{pic_mask-detection_synchro}
    }
    \caption[]{Scenario 4: Training set, test set, and predictions result for interpolation with 2 cores (red), 4 cores (blue), 8 cores (green), and 16 cores (brown) on each component.}\label{fig:mask_detection_synchro}
    \vspace{-5pt}
\end{figure}

We deploy the \textit{Mask detection} application on Amazon EC2 VMs, specifically using \textit{m4.large} instances\footnote{\scriptsize{\url{https://aws.amazon.com/ec2/instance-types/}}}.
Profiling data are collected with the same number of cores (2, 4, 8, and 16) for both the \textit{blurry-faces} and \textit{mask-detector} components.
Results for the ML performance models trained with this data are reported in Table \ref{table:MAPE-interpolation-scenarios_4_5}.
Figure \ref{fig:mask_detection_synchro} shows the interpolation predictions with the best models.
In this test campaign, the test set comprises profiling data originating from specific workloads (which are five evenly-spaced values for each number of cores).

The MAPE of the best combined model is 7.01\%.
Recall that in this case, we do not have to consider the pipeline effect, as synchronous calls represent a continuous flow of jobs, allowing our models to predict the average execution time of each individual job.
For this reason, the execution time of the full workflow is the sum of the times of the individual components.

\begin{table}[h!t]
\centering
\resizebox{\linewidth}{!}{
\begin{tabular}{ |p{3cm}|p{1cm}|p{2.5cm}|p{2.5cm}|p{2.5cm}|p{2.5cm}|p{2cm}|p{2cm}| }
\cline{3-8}
\multicolumn{2}{c}{} & \multicolumn{3}{|c|}{MAPE interpolation, \textit{Scenario 4}} & \multicolumn{3}{|c|}{MAPE interpolation, \textit{Scenario 5}}\\
\hline
Algorithm & SFS & Full workflow & blur-faces & mask-detector & Full workflow & partition 1 & partition 2  \\ \hline
\multirow{2}{*}{Ridge Regression} 
& Yes & 11.19 & 16.19 & 15.78 & 8.89 & 14.22 & 8.29 \\
\cline{2-8}
& No & \textbf{7.01} & 8.57 & 13.05 & 6.67 & 10.43 & \textbf{7.80} \\\hline
\multirow{2}{*}{Decision Tree} 
& Yes & 12.96 & 14.41 & 20.17 & \textbf{5.97} & 9.84 & 9.57 \\
\cline{2-8}
& No & 17.15 & 17.99 & 19.41 & 11.21 & 9.94 & 21.34 \\\hline
\multirow{2}{*}{XGBoost} 
& Yes & 28.60 & 44.55 & \textbf{12.55} & 9.57 & 12.02 & 12.66 \\
\cline{2-8}
& No & 9.23 & \textbf{6.09} & 15.87 & 10.32 & \textbf{9.64} & 16.46 \\\hline
\multirow{2}{*}{Random Forest} 
& Yes & 40.78 & 11.02 & 78.82 & 10.18 & 13.97 & 8.15 \\
\cline{2-8}
& No & 31.39 & 14.70 & 60.96 & 7.90 & 12.64 & 7.95 \\\hline
\end{tabular}
}
\caption{MAPE [\%] on interpolation in \textit{Scenario 4} and \textit{Scenario 5}.}\label{table:MAPE-interpolation-scenarios_4_5}
\vspace{-5pt}
\end{table}

\vspace{-5pt}
\paragraph{Scenario 5: impact of neural network partitioning}

In the last scenario, we demonstrate the capability of OSCAR-P to manage individual application components partitioned into multiple sub-components.
Specifically, we focus on the \textit{blurry-faces} component considering two partitions.
The YOLOv4 network is partitioned at \textit{layer 266}, corresponding to the layer that exports the smallest tensor.
As in the previous scenario, the components are deployed on \textit{m4.large} EC2 instances. 
Additionally, we explore the interpolation capability of ML models amidst variations in the number of cores and throughput.
Figure \ref{fig:blurryfaces_partitioned} shows the collected data and the corresponding predictions with the best models.
Table \ref{table:MAPE-interpolation-scenarios_4_5} shows the MAPEs for the interpolation tests of all models, with a best value of 5.97\%.

Note that the total execution time of the partitioned component in this particular setup exceeds that of the entire component mainly due to the transfer time between partitions.
Despite this, DNN partitioning is still relevant in practice to optimize resource usage, particularly on constrained devices \cite{tadakamalla2021autonomic}.

\vspace{-5pt}
\subsection{Results with Video searcher application}\label{sec:video_searcher_application}
The video searcher application, which includes six compute-intensive components (see Section~\ref{subsubsec:video-searcher-app}), is tested in the computing continuum. 
In particular, services are deployed on both local VM clusters, Raspberry Pi devices, and EC2 instances, as shown in Table \ref{table:setting grep big app} in \ref{app:exper-video}.
Notably, the full workflow is tested across a range of nodes for each component, from 1 to 8.

We perform asynchronous calls by uploading 1-minute-long videos in batches of 5, 10, and 15.
The DeepSpeech component is also tested individually to assess whether increasing the number of cores from 1 to 4 improves its performance.
We do observe an improvement in the test results, but it is not enough to justify running one component with 4 cores instead of 4 components with 1 core each.
One reason is that only the actual computation benefits from the increased number of cores, while other operations, such as the container setup, do not.
As in previous scenarios, we test both the full workflow and the single services.

First, we conduct an interpolation analysis including 16 and 24 cores in the test set. 
Subsequently, we use all the obtained values to generate the combined model to predict the runtime of the full workflow.
We show results for the best models obtained for data with 5, 10, and 15 input files in Figure \ref{fig:video-searcher-interpolation} in \ref{app:exper-video}: their MAPEs are 27.63\%, 18.95\%, and 7.88\%, respectively.

Next, we perform extrapolation on the number of input files.
In this case, the aMLLibrary training set is the collection of configurations with 5 or 10 files, while the test set includes the 15-video batch.
We also apply the pipeline effect correction explained in Section \ref{sec:mask_detection_application_async}.
Figure \ref{fig:video-searcher-extrapolation} in \ref{app:exper-video} shows the training set, the test set, and the extrapolation prediction.
Even though we are considering an application with many services distributed across three clusters and training our models with only two different input sizes, we obtain a MAPE of 22.33\% for the full workflow.
This outcome is especially remarkable given the constraints of our testing campaign.

\vspace{-5pt}
\subsection{Results with Recipe transcriber application}\label{sec:recipe_transcriber_application}
The recipe transcriber application is tested using both Amazon EC2 instances and AWS Lambda\footnote{\scriptsize{\url{https://aws.amazon.com/lambda/?nc1=h_ls}}}, which is one of the main platforms providing FaaS.
This is the most complex scenario we consider, as the application consists of seven compute-intensive components (see Section \ref{subsubsec:recipe-transcriber-app}).
In the previous section, we analyze the first five components.
Here, we focus on the final two components, deployed on AWS Lambda, and on estimating the execution time of the entire pipeline.
We perform asynchronous calls using a batch of 100 videos, each with a duration of 10 seconds.
In this scenario, we observe the ability of our performance models to predict the average execution time of the entire pipeline for each individual video input.

Note that for such a large batch of video inputs, it is crucial to consider that the arrival rate of each component depends on the output rate of previous components, which in turn hinges on exogenous load and component configurations.

Furthermore, unlike previous campaigns, we also assign different numbers of cores to each component.  
Table \ref{table:setting big app Yolo V4} in \ref{app:exper-recipe} displays all possible configurations of the individual components, listing the number of nodes, cores, and the level of parallelism, i.e., the number of parallel instances per component.

Note that the EC2 instances are selected to ensure that two pods are instantiated on each node.
Furthermore, for the \textit{ffmpeg-2} component, the number of cores is larger than (specifically, a multiple of) the parallelism level due to the high computational load of the component.
However, the parallelism level of Lambdas is undefined, since AWS Lambda guarantees automatic resource scalability and therefore jobs are handled with full parallelism by the platform with no limitations.

This variety of component configurations results in a total of $2 \times 3 \times 2 \times 4 \times 2 = 96$ possible configurations for the full workflow, a number that would increase exponentially if we tested more than two different configurations for all components.
In general, testing all possible configurations would be unfeasible in terms of time and budget.
Thus, we devise a strategy to limit the number of profiled setups to one third of the total 
by only considering the profiling configurations with the smallest and largest \textit{parallelism levels} for each component.
Therefore, the training set for our models consists of the $2^5 = 32$ combinations of configurations (see Table \ref{table:setting big app Yolo V4} in \ref{app:exper-recipe}).

Moreover, for each component after the first one, we included in the training set only profiling data collected when considering the lowest parallelism level on all the previous components.
This process guarantees an optimal balance between opposing scenarios: on one hand, we consider cases where we have performance dependency in prior components, and, on the other hand, cases in which files arrive in batches due to the large input parallelism.
All the other combinations constitute the test set.

Since all intermediate configurations are excluded from the training set, linear regression is the most suitable model for estimating runtime.
As we only use training data at the extremes of the intervals, models such as Random Forest and Decision Tree would not be able to discriminate between values within those intervals\footnote{Random Forest and Decision Tree provide stepwise regression functions. Hence, intermediate configurations would be approximated with the values of the minimum and maximum core configurations.}.
Thus, we train a Ridge Regression model for each component without SFS, estimate their execution times on the test set, and evaluate the MAPE on the total time. 
The MAPE on the test set for the \textit{ffmpeg-3} and \textit{object detector} components are 5.01\% and 2.48\%, respectively, while the MAPE for the entire workflow is 26.78\%.
This result is noteworthy especially given the significant inherent variability present in the profiling data and the fact that our models do not consider the dependency of the component execution time on the parallelism of all preceding components.

\vspace{-5pt}
\subsection{Results with Fibonacci application}\label{sec:fibonacci_application}
We test OSCAR-P on the Fibonacci application by relying on synchronous calls.
In this scenario, we test an application with input files of few bytes to examine whether the trend in execution times differs from previous applications with larger inputs.
Specifically, we deploy the unique component on a physical cluster consisting of 7 VMs, each provisioned with 4 GB of memory and 4 vCPUs.
As outlined in Section \ref{sec:mask_detection_application_sync}, we test the application with varying numbers of cores and input workloads.
In the 4-core setting, we fix the input workload at $\lambda \in [0.2, 0.6]$ req/s, with the upper bound representing the point at which we observe saturation.
As the number of cores increases, we gradually raise this upper bound, up to 3.5 req/s in the 28-core case.
We force this constant workload for 10 minutes via JMeter, with half of the cores being warm at the beginning of the test.

We report the results of the interpolation tests with ML models in Table \ref{table:MAPE-interpolation-fibonacci}.
Also, Figure \ref{fig-fibo-app} in \ref{app:exper-fibonacci} displays the training and test sets along with the prediction obtained using the best model.
\begin{table}[h!t]
\centering
\resizebox{5cm}{!}{
    \begin{tabular}{ |p{3cm}|p{1cm}|p{3cm}| }
    \hline
    Algorithm & SFS & Full workflow \\\hline
    \multirow{2}{*}{Ridge Regression}
    & Yes & 15.24 \\
    \cline{2-3}
    & No & 13.15 \\\hline
    \multirow{2}{*}{Decision Tree}
    & Yes & 11.45 \\
    \cline{2-3}
    & No & 11.87 \\\hline
    \multirow{2}{*}{XGBoost}
    & Yes & 10.98 \\
    \cline{2-3}
    & No & 10.15 \\\hline
    \multirow{2}{*}{Random Forest}
    & Yes & \textbf{9.64} \\
    \cline{2-3}
    & No & 9.91 \\\hline
    \end{tabular}
}
\caption{MAPE [\%] on interpolation.}\label{table:MAPE-interpolation-fibonacci}
\vspace{-5pt}
\end{table}
These results indicate that the ML models achieve errors below 10\%, showing the remarkable predicting capabilities of OSCAR-P even in the case of small inputs and large workloads.

\vspace{-5pt}
\subsection{Discussion}  \label{subsec:exper-discussion}
Using the OSCAR-P framework for application profiling brings notable efficiency in time and budget allocation for conducting performance profiling campaigns.
OSCAR-P automates and optimizes profiling, eliminating all periods of inactivity associated with manual profiling, during which machines remain idle and logs are downloaded manually.
We provide an example of savings related to the ML performance models interpolation capabilities by considering the most complex application analyzed in this work, namely \textit{Recipe transcriber} presented in Section \ref{subsubsec:recipe-transcriber-app}.
In that case, the collection of profiling data for all possible combinations of input configurations took about two weeks and had a total cost of about 200 USD.
If we only consider executions associated with configurations in the training set of the ML models, the data collection time decreases slightly over five days.
The performance of the remaining configurations (i.e., the ones in the test set) can be estimated with acceptable accuracy by the trained ML models, therefore in a real use case it would not be necessary to profile them directly, reducing the costs by about 55\%.
The reduction would have been even more significant if we had considered larger levels of parallelism for each component.
\RStext{Overall, we have shown that by profiling only 33\% of all possible configurations, we can obtain a model with a prediction error of less than 30\%.}

\vspace{-8pt}
\section{Conclusions and future work}\label{sec:conclusions}
\vspace{-5pt}

This work introduces OSCAR-P, an auto-profiling framework, and aMLLibrary, an autoML solution for performance prediction.
These tools automate the testing of application workflows on various hardware configurations and generate ML-based performance models.
OSCAR-P significantly reduces setup time and manual processing by automating the execution of profiling experiments and data collection.
Additionally, aMLLibrary enables accurate performance model training without ML expertise.
\RStext{Experimental results show that our models achieve a MAPE smaller than 30\%, which makes them useful for design-time decisions and runtime resource management.

Future work will primarily address certain limitations of our study.
First, we plan to test a broader range of hardware specifications for the same application.
In particular, including the machine specifications (besides the number of cores) as input features would allow us to generalize our performance models across different setups.
We aim to examine whether the performance of aMLLibrary is affected and if it can capture common trends across different kinds of machines.

Second, in this work, we attempted to minimize the network effect to focus on computational performance.
However, real-world environments often face network-related issues such as congestions or jitters.
A potential research direction would be to analyze how the network influences throughput.
We expect that the network effect mainly translates into reduced input rates for the application components.
Since our models for individual components also account for throughput, we will evaluate if they still remain effective in such conditions.

Finally, we intend to scale our tools to an industrial level.
The most computationally intensive application we evaluated in this study, \textit{Recipe Transcriber}, involved testing  configurations using up to 82 total cores on AWS and  Lambda functions.
Our next step is to validate our models on applications including tens of components spanning across hundreds or thousands of nodes.
We expect increased variance at this scale due to more severe resource contention and amplification of performance noise across component dependencies.
}

\vspace{-8pt}
\section*{Acknowledgment}
\vspace{-9pt}
\begin{footnotesize}
\noindent The European Commission funded this work under the H2020 project AI-SPRINT (grant n. 101016577).
OSCAR was developed under project PDC2021-120844-I00 funded by MCIN/AEI/-10.13039/5011000-11033 and by the European Union NextGenerationEU/PRTR and under grant PID2020-113126RB-I00 funded by MCIN/AEI/10.13039/501100011033.
aMLLibrary was developed under the H2020 HPC JU LIGATE (grant n. 956137). 
\end{footnotesize}

\vspace{-5pt}
\begin{footnotesize}
\setstretch{0.25}
\bibliography{00_biblio}
\end{footnotesize}

\newpage

\appendix
\section{Supplemental material}
In this section, we provide some additional material that includes details about the configuration of aMLLibrary in \ref{app:aml-config} and additional plots and analyses from our experimental campaign in \ref{app:exper-mask} through \ref{app:exper-fibonacci}.

\subsection{aMLLibrary configuration file}  \label{app:aml-config}
We report the example configuration file for aMLLibrary mentioned in Section \ref{sec:aMLLib}.
Note that this file represents an equivalent Python script of hundreds of lines of code (see the equivalent script to the Appendix configuration file in the Zenodo repository \cite{zenodo_link}).

\begin{multicols}{2}
\begin{singlespacing}
\footnotesize{
\begin{verbatim}
[General]
techniques = ['LRRidge', 'XGBoost',
              'DecisionTree']
hp_selection = KFold
validation = HoldOut
folds = 5
hold_out_ratio = 0.2
y = exec_time
hyperparameter_tuning = Hyperopt
hyperopt_max_evals = 10

[DataPreparation]
input_path = dataset.csv
normalization = True
product_max_degree = 2
inverse = [*]

[LRRidge]
alpha = ['loguniform(0.01,1)']
[XGBoost]
min_child_weight = [1]
gamma = ['loguniform(0.1,10)']
n_estimators = [500]
learning_rate = ['loguniform(0.01,1)']
max_depth = [10]
alpha = [0]
lambda = [1]

[DecisionTree]
criterion = ['squared_error']
max_depth = [3]
max_features = ['auto']
min_samples_split = ['loguniform(0.01,1)']
min_samples_leaf = ['loguniform(0.01,0.5)']



\end{verbatim}}
\end{singlespacing}
\end{multicols}

\subsection{aMLLibrary hyperparameters}  \label{app:aml-hypers}
We list the values or prior distributions of ML models trained by aMLLibrary in our experimental campaigns in Table \ref{tab:oscarp-hyperparameters}.
We recall that BO tuning samples a maximum of 10 parameter sets for each model.
\begin{table}[h!t]
\centering
\resizebox{7cm}{!}{
\begin{tabular}{ |p{3.0cm} p{4.5cm} p{3.0cm}| }
\hline
\textbf{Algorithm} & \textbf{Hyperparameter Name} & \textbf{Values}\\ \hline
\multirow{1}{*}{Ridge Regression}
& alpha & loguniform(0.01,1) \\\hline
\multirow{5}{*}{XGBoost}
& min\_child\_weight & 1 \\
& gamma & loguniform(0.1,10) \\
& n\_estimators & 1000 \\
& learning\_rate & loguniform(0.01,1) \\
& max\_depth & 100 \\\hline
\multirow{5}{*}{Decision Tree}
& criterion & mse \\
& max\_depth & 3 \\
& max\_features & auto \\
& min\_samples\_split & loguniform(0.01,1) \\
& min\_samples\_leaf & loguniform(0.01,0.5) \\\hline
\multirow{5}{*}{Random Forest}
& n\_estimators & 5 \\
& criterion & mse \\
& max\_depth & quniform(3,6,1) \\
& max\_features & auto \\
& min\_samples\_split & loguniform(0.1,1) \\
& min\_samples\_leaf & 1 \\\hline
\end{tabular}
}
\caption{Hyperparameters used for each ML algorithm.}\label{tab:oscarp-hyperparameters}
\end{table}

\subsection{Mask detection experimental results}  \label{app:exper-mask}
We depict the Figures related to the results with the Mask detection application, presented in Section \ref{sec:mask_detection_application}.

\begin{figure}[h!t]
    \centering
    \subfloat[Runtimes of the computation phases]{
        \includegraphics[width=0.48\textwidth]{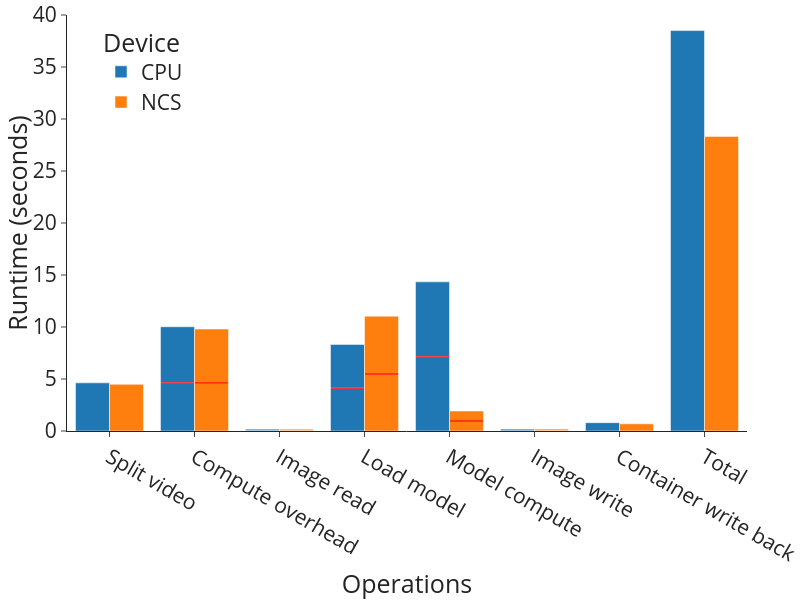}\label{fig:ncs-pi-bars}
    }
    \subfloat[Total runtime over number of input files]{
        \includegraphics[width=0.48\textwidth]{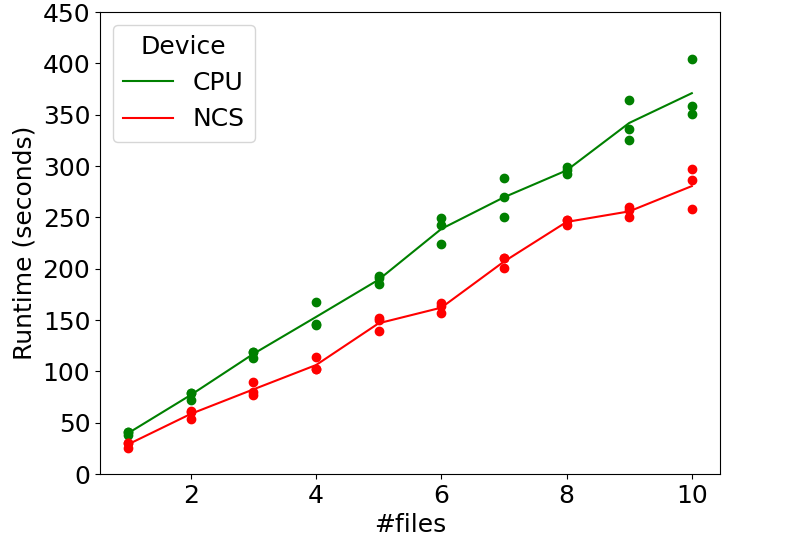}\label{fig:ncs-pi-lines}
    }
    \caption[]{Scenario 3: Comparison of the operations runtime with and without accelerator.}\label{fig:ncs-pi-comparison}
\end{figure}

Figure \ref{fig:ncs-pi-comparison} shows the results of executions with and without the NCS2 described in Section \ref{sec:mask_detection_application_async}, \textit{Scenario 3}.
In particular, in Figure \ref{fig:ncs-pi-bars}, some bars are split in half by red horizontal lines, representing each of the two produced frames for each input video. 

\begin{figure}[h!t]
    \centering
    \begin{minipage}[]{.32\linewidth}
        \includegraphics[width=\textwidth]{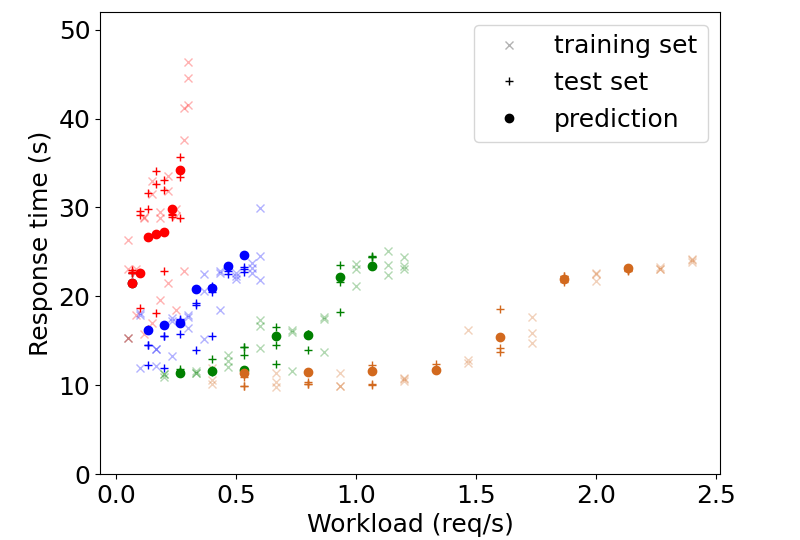}
        \subcaption{Full workflow}
        \label{fig:blurryfaces_complete}
    \end{minipage}
    \hfill
    \begin{minipage}[]{.32\linewidth}
        \includegraphics[width=\textwidth]{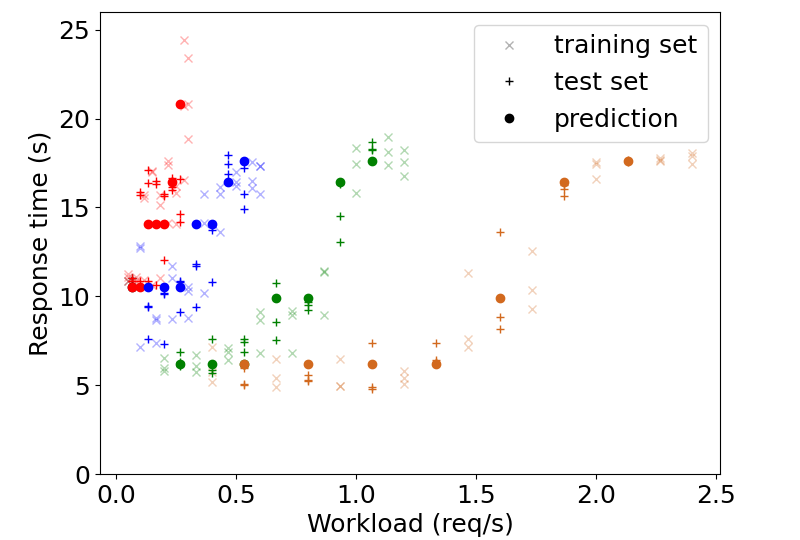}
        \subcaption{partition 1}
        \label{fig:blurryfaces_partition1}
    \end{minipage}
    \hfill
    \begin{minipage}[]{.32\linewidth}
        \includegraphics[width=\textwidth]{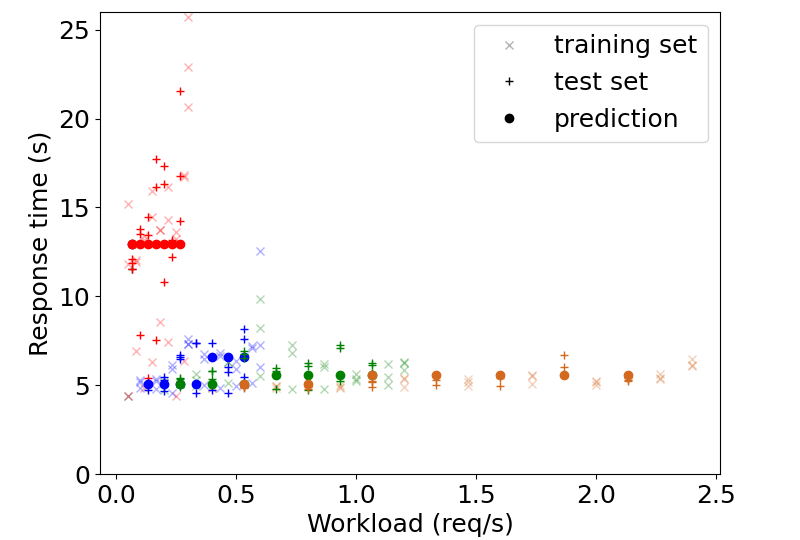}
        \subcaption{partition 2}
        \label{fig:blurryfaces_partition2}
    \end{minipage}
    \caption[]{Training set, test set, and predictions result for interpolation with 2 cores (red), 4 cores (blue), 8 cores (green) and 16 cores (brown) on each partition of the blur-faces component.}\label{fig:blurryfaces_partitioned}
\end{figure}

Figure \ref{fig:blurryfaces_partitioned} shows the profiling data, divided into training and test sets, along with the predictions for the full workflow of the experimental campaign on the \textit{blurry-faces} component partitioned as described in Section \ref{sec:mask_detection_application_sync}, Scenario 5. 

\subsection{Video searcher application experimental results} \label{app:exper-video}
We show Figures and Tables related to the results with the Video searcher application, presented in Section \ref{sec:video_searcher_application}.

Table \ref{table:setting grep big app} reports the application settings, including the type of computational resource, the number of nodes, and the corresponding number of cores for each component.

\begin{table}[h!t]
\centering
\resizebox{6cm}{!}{
\begin{tabular}{ |c|c|c|c| } 
\hline
\textbf{Service} & \textbf{Resource} & \textbf{\#Nodes} & \textbf{\#Cores} \\ \hline
ffmpeg-0 & local VMs & \multirow{6}{*}{$1,2,\dots,8$} & \multirow{6}{*}{$4,8,\dots,32$} \\
librosa & local VMs & & \\
ffmpeg-1 & Raspberry Pi & & \\
ffmpeg-2 & local VMs & & \\
DeepSpeech & AWS t2.xlarge & & \\
Grep & AWS t2.xlarge & & \\\hline 
\end{tabular}
}
\caption{Setting of the simulations of video searcher application.}\label{table:setting grep big app}
\end{table}

\begin{figure}[h!t]
    \centering
    \begin{minipage}[]{.32\linewidth}
        \includegraphics[width=\textwidth]{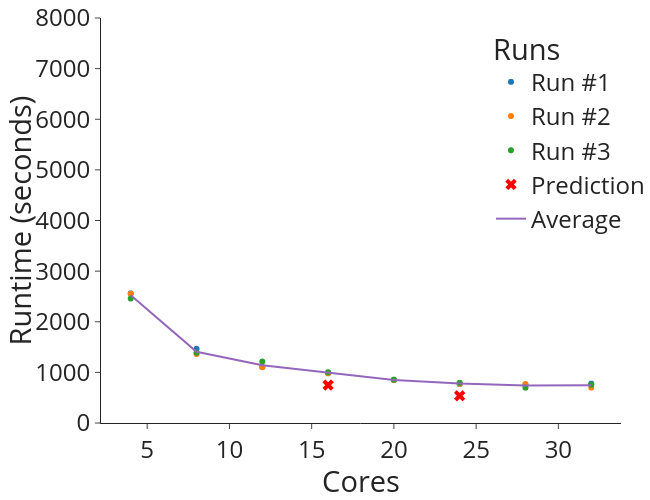}
        \subcaption{5 files}
    \end{minipage}
    \hfill
    \begin{minipage}[]{.32\linewidth}
        \includegraphics[width=\textwidth]{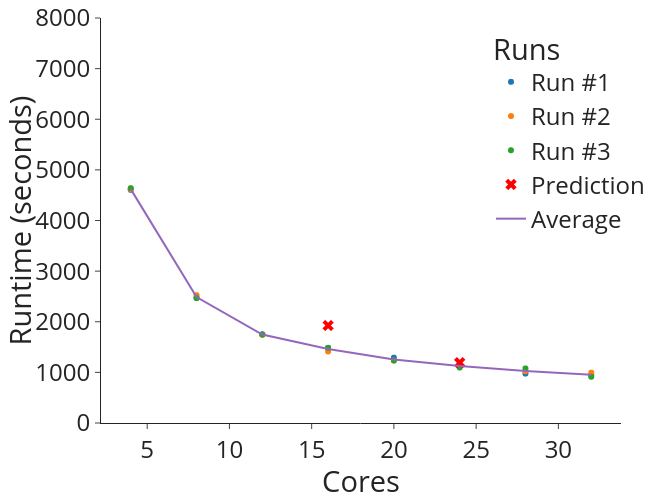}
        \subcaption{10 files}
    \end{minipage}
    \hfill
    \begin{minipage}[]{.32\linewidth}
        \includegraphics[width=\textwidth]{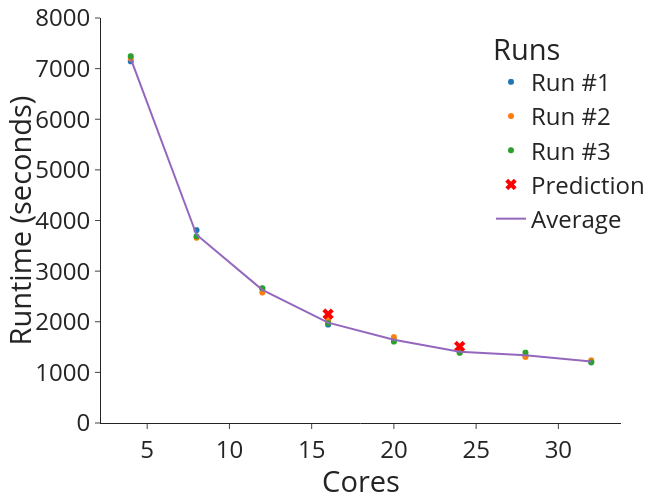}
        \subcaption{15 files}
    \end{minipage}
    \caption[]{Interpolation tests for the ``Video searcher'' application using combined model.}\label{fig:video-searcher-interpolation}
\end{figure}

\begin{figure}[h!t]
    \centering
    \subfloat[Full workflow]{
        \includegraphics[width=0.4\textwidth]{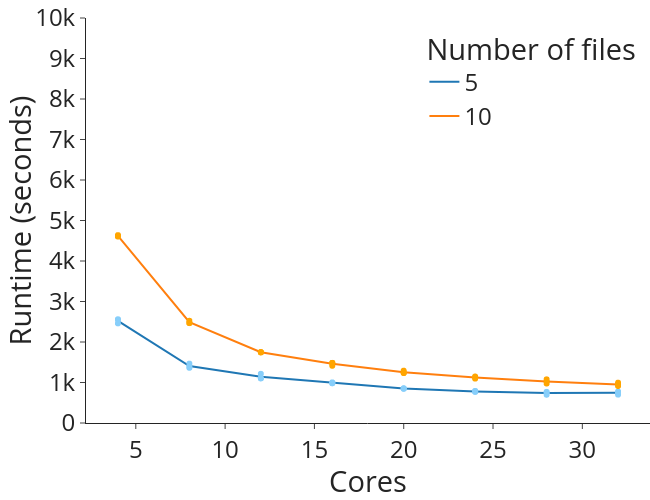}
    }
    \subfloat[Test set and prediction on the combined model]{
        \includegraphics[width=0.4\textwidth]{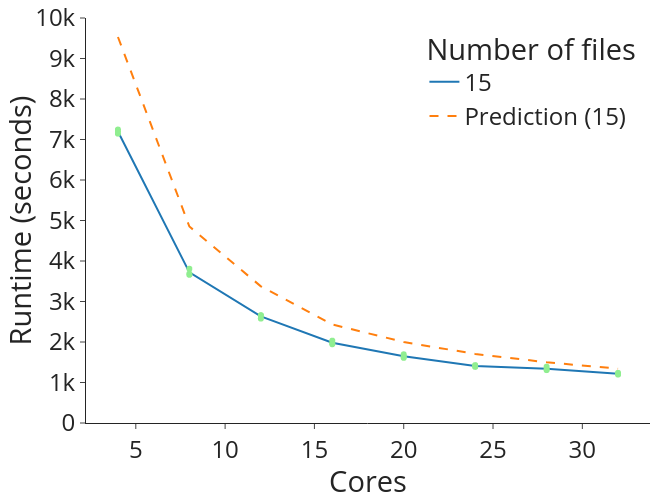}
    }
    \caption{Extrapolation tests for ``video-searcher'' application.}\label{fig:video-searcher-extrapolation}
\end{figure}

Figures \ref{fig:video-searcher-interpolation} and \ref{fig:video-searcher-extrapolation} show the results of the interpolation and extrapolation tests on data collected with OSCAR-P.

\subsection{Recipe transcriber application experimental results}  \label{app:exper-recipe}
Table \ref{table:setting big app Yolo V4} shows the settings for the experimental results with the Recipe transcriber application, described in Section \ref{sec:recipe_transcriber_application}, including the type of computational resource, the number of nodes, and the corresponding number of cores for each component.

\begin{table}[h!t]
\centering
\resizebox{8cm}{!}{
\begin{tabular}{ |c|c|c|c|c| } 
\hline
\textbf{Component} & \textbf{AWS instance} & \textbf{Parallelism level} & \textbf{\#Cores} & \textbf{\#Nodes} \\ \hline
ffmpeg-0 & m4.large & \textbf{2}, \textbf{4} & 2, 4 & 1, 2 \\ \hline
librosa & m4.large & \textbf{2}, 4, \textbf{6} & 2, 4, 6 & 1, 2, 3 \\ \hline
ffmpeg-1 & m4.large & \textbf{2}, \textbf{4} & 2, 4 & 1, 2 \\ \hline
ffmpeg-2 & m4.4xlarge & \textbf{2}, 4, 6, \textbf{8} & 16, 32, 48, 64 & 1, 2, 3, 4 \\ \hline
DeepSpeech & m4.large & \textbf{2}, \textbf{4} & 2, 4 & 1, 2 \\ \hline
ffmpeg-3 & Lambda & N/A & N/A & N/A \\ \hline
object detector & Lambda & N/A & N/A & N/A \\ \hline
\end{tabular}
}
\caption{Setting of recipe transcriber application. We highlight in bold the levels of parallelism used for the combinations of configurations in the training set.}\label{table:setting big app Yolo V4}
\end{table}

\subsection{Fibonacci application experimental results}  \label{app:exper-fibonacci}
\begin{figure}[ht]
    \centering
    \includegraphics[width=0.6\textwidth]{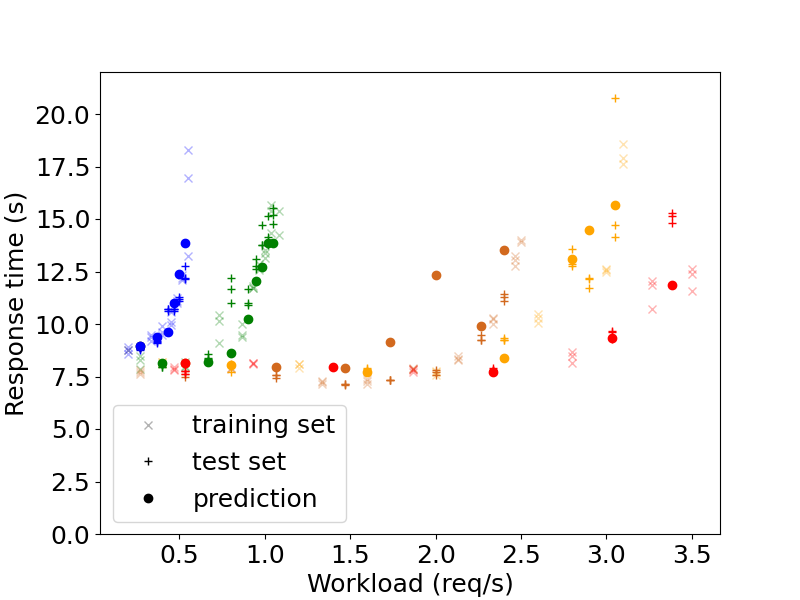}
    \caption{Training set, test set, and predictions result for interpolation with 4 cores (blue), 8 cores (green), 16 cores (brown), 24 cores (orange) and 28 cores (red) on Fibonacci application.}\label{fig-fibo-app}
\end{figure}

Figure \ref{fig-fibo-app} shows the profiling data, divided into training and test sets, along with the predictions for the Fibonacci application analysis, as described in Section \ref{sec:fibonacci_application}.

\end{document}